\documentclass[fleqn,usenatbib]{mnras}

\usepackage{newtxtext,newtxmath}

\usepackage[T1]{fontenc}

\DeclareRobustCommand{\VAN}[3]{#2}
\let\VANthebibliography\thebibliography
\def\thebibliography{\DeclareRobustCommand{\VAN}[3]{##3}\VANthebibliography}

\usepackage{graphicx}	
\usepackage{amsmath}	

\usepackage[table,xcdraw]{xcolor}
\usepackage[font=small,skip=0pt]{caption}
\usepackage{orcidlink}
\usepackage{booktabs}
\usepackage{lscape}
\usepackage{url}
\usepackage{stfloats}
\usepackage{lineno}
\usepackage{caption}
\usepackage{subcaption}
\usepackage{soul}

\renewcommand{\d}[1]{\ensuremath{\operatorname{d}\!{#1}}}

\newcommand{\pkg}[1]{\texttt{#1}}

\title[Modelling of Astrophysical Systematics]{Joint Modelling of Astrophysical Systematics for Cosmology with LSST Cosmic Shear
}

\author[N. \v{S}ar\v{c}evi\'c et al.]{
Nikolina \v{S}ar\v{c}evi\'c,$^{1}$
\orcidlink{0000-0001-7301-6415}
\thanks{E-mail: n.sarcevic2@newcastle.ac.uk, nikolina.sarcevic@gmail.com},
C. Danielle Leonard$^{1}$
\orcidlink{0000-0002-7810-6134},
Markus M. Rau$^{1,2}$
\orcidlink{0000-0003-3709-1324},
\newauthor
and the LSST Dark Energy Science Collaboration
\\
$^{1}$School of Mathematics, Statistics and Physics, Newcastle University, Herschel Building, NE1 7RU Newcastle-upon-Tyne, UK\\
$^{2}$High Energy Physics Division, Argonne National Laboratory, Lemont, IL 60439, USA\\
}

\date{Accepted XXX. Received YYY; in original form ZZZ}

\pubyear{2024}

\begin{document}
\label{firstpage}
\pagerange{\pageref{firstpage}--\pageref{lastpage}}
\maketitle

\begin{abstract}
We present a novel framework for jointly modelling the weak lensing source galaxy redshift distribution and the intrinsic alignment (IA) of galaxies through a shared luminosity function (LF). In the context of a Rubin Observatory's Legacy Survey of Space and Time (LSST) Year 1 and Year 10 cosmic shear analysis, we show that our novel approach produces cosmological parameter constraints which are comparable to standard methods, while offering more physical insight into IA and selection effects.
We clarify the relationship between individual parameters of a Schechter luminosity function and the redshift distribution of a magnitude-limited sample, showing the consequences of marginalizing over these parameters when modelling intrinsic alignments in standard cosmic shear analyses.
We explore the impact of the shape of the luminosity function on the cosmic shear data vector, and we outline the potential of this method to naturally model selection functions in redshift distribution estimation.
Although this work focuses on LSST cosmic shear, the proposed joint modelling framework is broadly applicable to weak lensing surveys.

\end{abstract}




\section{Introduction}
\label{sec:intro}
Weak lensing (WL) is a powerful technique used in cosmology to probe the properties of the universe with high precision. 
Several Stage III surveys, including the Kilo Degree Survey \citep{KiDSDR4}, the Dark Energy Survey \citep{DESDR2}, and Hyper Suprime-Cam \citep{HSCDR3}, have already demonstrated its effectiveness in this regard, with constraints on specific cosmological parameters comparable to those from the Cosmic Microwave Background (CMB).
However, upcoming Stage IV surveys such as the Rubin Observatory's Legacy Survey of Space and Time (LSST) \citep
{Ivezic2019}, {\it Euclid} \citep{Euclid2022, EuclidDef}, and the Roman Space Telescope \citep{eifler2021cosmology} promise to further improve the precision of weak lensing measurements.
To harness the full potential of these future surveys, it is crucial to develop accurate models for their observables from which we will extract cosmological parameter constraints. 
An integral component of this task is the modelling of non-cosmological effects, as any inaccuracies in modelling could lead to biases in the cosmological parameter estimation.

Among the significant challenges in modelling non-cosmological effects on weak lensing are astrophysical systematic effects, including the impact of baryonic physics, the galaxy-dark matter bias (for cross-correlations between lensing and galaxy clustering), and, as will be a focus of this work, the intrinsic alignment of galaxies.
Intrinsic alignment (IA) refers to the scenario where galaxies exhibit correlated alignment not due to the gravitational lensing of their light but as a result of local effects, predominantly tidal physics as well as galaxy evolution history and environment \citep{Joachimi_2015, Kirk_2015, Troxel_2015}. 
If ignored or incorrectly modelled, the IA impact to the weak lensing cosmological signal has the potential to severely bias cosmological parameter inference \citep{Krause_2015, secco2022dark}. 
Extensive efforts to understand and model intrinsic alignment have thus become a priority, with direct measurement using spectroscopic samples \citep{mandelbaum2011wigglez, joachimi2011constraints, singh2015intrinsic, kids+gama2019}, the analysis of alignments in hydrodynamical simulations \citep{Chisari2015, tenneti2016intrinsic, samuroff2021advances} and theoretical modelling \citep{hirata2004intrinsic, Bridle_2007, blazek2015tidal, vlah_eft_202, hymalaia_2023arXiv230713754M, bakx_eft2023JCAP...10..005B, lagrangian_shapes2024JCAP...01..027C} all playing a part.

Besides intrinsic alignment modelling, large area photometric survey programs for weak lensing cosmology need to model the redshift distribution of samples of galaxies to unprecedented precision \citep{srd_lsst_desc}.
Of particular importance is the accurate and precise quantification of the photometric redshift (PZ) distribution uncertainty, which will require estimation and modelling methodology going beyond that which has been employed in Stage III analyses (see \citealt{rau2023weak, bilicki2021bright, myles2021dark} for examples of Stage III photometric redshift methodologies and \citealt{zhang2023photometric, moskowitz2023improved, rau2020estimating} for examples of new techniques targeted at Stage IV).

The standard technique for mitigating the impact of both IA and PZ uncertainty involves including a model for these effects in the theoretical calculation for weak lensing observables, as part of the parameter inference framework, and marginalizing over the parameters of these models when quoting cosmological results (see, \textit{e.g.} \citealt{des_year3, hsc_year3, kids1000_cataloge}). Currently, these state-of-the-art approaches treat IA and PZ modelling as independent of each other. 
However, in recent years it has become increasingly recognized that the modelling of intrinsic alignment and photometric redshift distributions within cosmic shear analyses are not fully isolated from each other, with suggestions of trade-offs and interplay between the parameters of the two effects \citep{wright2020kids, stolzner2021self, fischbacher2023redshift}.
\citet{leonard2024photometric} recently explored this in the case of a 3$\times$2pt analysis for an LSST-like survey, finding that although biases to cosmological parameters due to model mis-specification of PZ could often be mitigated by marginalization over a sufficiently flexible IA model, this was not always the case and in general led to significant biases in the IA and PZ parameters themselves.
The assumption that IA and PZ uncertainty are independent thus may leave potentially valuable information on the proverbial systematics table.

In this work, we propose a method to link the modelling of IA and PZ uncertainty for cosmic shear, via their mutual dependence on the luminosity of galaxies in the weak lensing source sample.
We know that the ensemble redshift distribution in a magnitude-limited sample can be modelled as an integral over a selection function and a luminosity function (see \textit{e.g.} \citealt{van_Daalen_2018}). 
Equally, astrophysical systematic effects including IA  \citep{blazek2011testing, Chisari2015, Fortuna_2020, samuroff2023dark}, as well as galaxy bias have been shown to be galaxy-luminosity dependent.
As shown in \citet{van_Daalen_2018} for the case of galaxy clustering and luminosity-dependent galaxy bias, taking advantage of the link between luminosity-dependent astrophysical systematic effects and luminosity-based modelling of redshift distributions in magnitude-limited surveys can provide a natural mechanism for extracting valuable, degeneracy-breaking information.

In this work, we propose a novel modelling framework that exploits the physical connection between the intrinsic alignments and the redshift distribution of galaxies in a magnitude-limited survey. 
Extending the work of \citet{joachimi2011constraints}, \citet{Krause_2015}, and \citet{van_Daalen_2018}, our approach uses a shared luminosity function to simultaneously model the intrinsic alignment amplitude and the redshift distribution for a magnitude-limited sample.
By adopting a shared luminosity function, we provide a method for self-consistently constraining luminosity function parameters (which are typically fixed based on external datasets if included at all), while also introducing a framework for modelling generic selection functions (which often arise when estimating sample redshift distributions, e.g. in calibration with spatially overlapping spectroscopic data).
Although in this initial proof-of-concept work we incorporate photometric redshift information only via the impact of the redshift distribution on the weak lensing two-point datavector, our method lays the groundwork for a more general unified framework to incorporate a single luminosity function within both modelling of astrophysical systematics and estimation of redshift distributions via spectral fitting and cross-correlation methods.

While the focus of this work is primarily on cosmic shear, which quantifies the two-point correlations in the distortion of galaxy shapes due to weak gravitational lensing, the proposed modelling framework has the potential to be extended to 3$\times$2pt analysis and thus incorporate also the luminosity dependence of galaxy bias. 
The latter involves combining cosmic shear with galaxy-galaxy lensing and galaxy clustering measurements to extract even more valuable cosmological information from the data.

This work is structured as follows: Section \ref{sec:theory_and_methods} provides the theoretical description of key analysis ingredients: the luminosity function, redshift distribution, intrinsic alignments, and the cosmic shear power spectrum. 
Section \ref{sec:forecasting_setup} describes our forecasting setup.
Section \ref{sec:results} presents the results of an exploration of the potential of this method using Fisher forecasting. We conclude in Section \ref{sec:conclusions}.
\section{Theory and Methods}
\label{sec:theory_and_methods}

\subsection{Luminosity function of galaxies}
\label{subsec:luminosity_function}

The galaxy luminosity function $\phi (L) \d L$ represents the volume density of galaxies within a luminosity interval from $L$ to $L + \d L$.
The luminosity function (hereafter denoted LF) is a crucial and fundamental statistic used to describe galaxy populations and their evolution.
A fairly simple model of the number of galaxies in a luminosity interval is given by a classic Schechter function \citep{1976ApJ...203..297S}, in which the luminosity function scales as both a power law and exponentially:
\begin{equation} \label{eq:schechter_luminosity_function}
    \phi (L) \d L = \phi^* \left( \frac{L}{L^*} \right)^\alpha \exp \left( - \frac{L}{L^*} \right) \frac{\d L}{L^*}.
\end{equation}
Here, $\phi^*$ is a normalization factor, $\alpha$ is a power law index, commonly called a faint-end slope, while  $L^*$ represents a pivot luminosity --- a turning point at which the function starts to exponentially decrease from a power law.
Since we are interested in the redshift, $z$, scaling of the LF, the Schechter function can be expressed as:
\begin{equation} \label{eq:luminosity_function_z_scaling}
    \phi (L, z)dL  = \phi^* (z) \left( \frac{L}{L^*(z)} \right)^\alpha \exp \left( - \frac{L}{L^* (z)} \right) \frac{\d L}{L^* (z)}.
\end{equation}
Considering that it is often more convenient to work with  magnitudes rather than luminosities due to their closer connection to observables, and using the connection between the luminosities and magnitudes $\phi (L) \d L = \Phi (M) \d M$ \citep{Longair:2008gba}, the standard representation of the Schechter LF in the absolute magnitude ($M$) space used in this work is:
\begin{equation}
\label{eq:luminosity_function_magnitude_scaling}
\begin{split}
    \Phi (M, z) \d M &= 0.4 \: \phi^*(z) \ln(10) \left(10^{0.4(M - M^*(z))} \right)^{\alpha + 1} \\ &\times \exp \left( -10^{0.4(M - M^*(z))}\right) \d M, 
    \end{split}
\end{equation}
with $M^*(z)$ a characteristic magnitude (corresponding to the pivot luminosity).
We have adopted the classic but still widely used parametrization of the LF parameters from \citet{Lin_1999} (see e.g. \citealt{Loveday_2011}, \citealt{farrow2015galaxy}, and \citealt{karademir2023measurement} for examples of usage). Under this choice, the redshift evolution of the Schechter parameters is given by:

\begin{equation} \label{eq:lf_params}
\begin{split}
\phi^* (z) & = \phi_0 ^* 10^{0.4 P z}, \\
M^* (z) & = M^*_0 |_{z_0} - (z - z_0)Q, \\
\alpha (z) & = \alpha |_{z_0}.
\end{split}
\end{equation}

In this scheme, $Q$ is a scalar that represents a luminosity evolution parameter, describing a linear redshift evolution of characteristic absolute magnitude $M^*$. 
The faint-end slope parameter $\alpha$ is taken to be fixed in redshift in order for the LF shape to be constant.
Lastly, $z_0$ is a fixed pivot redshift.
An important consequence of this choice is that in this case, the normalization factor  $\phi^*(z)$ is closely related to the total galaxy number density $\rho = \int \phi (M) \d M$.
This way, the scaling of the total number density of galaxies is regulated by the density evolution parameter $P$, as described in Eq. \ref{eq:lf_params}.
Furthermore, the redshift evolution of the luminosity density $\rho_L = \int L \phi (M) \d M$ is then conveniently described by evolution parameters $P$ and $Q$:
\begin{equation} \label{eq:luminosity_density}
    \rho_L = \rho_L (0) 10^{0.4 (P + Q) z}.
\end{equation}
In constructing the fiducial values of the LF parameters used in this work, we make use as a starting point the parameters from the GAMA survey \citep{Loveday_2011}.
The GAMA survey based their LF parameter estimation on the maximum-likelihood approach of \citet{Lin_1999} for the CNOC2 galaxy redshift survey, but their results differ from those of CNOC2 because of differences in the two samples.
We have opted for the GAMA results because of deeper redshift coverage, the completeness in terms of passband filters, and clear distinction between the galaxy types.
In this regard, the starting point luminosity-function parameters of our analysis will be $r$-band LF parameters for red and all (red+blue) galaxies (Table \ref{tab:GAMADEEP2_lf_params}) with the pivot redshift $z_0 = 0.1$.
Details of how we adjust these parameters to find appropriate fiducial luminosity function parameters for our setup are discussed below.

We emphasize that, although in this work we use a Schechter LF parametrization described above, the method which we introduce and explore can be applied with any parametrization of the luminosity function.

\begin{table*}
\centering
\caption{The luminosity function parameters used as a starting point in the modelling.
Following \citet{Krause_2015, cosmolike_Krause_2017, srd_lsst_desc}, parameters are a combination of $r$-band fit parameters from GAMA survey \citep{Loveday_2011}, $B$-band parameters from DEEP2 survey \citep{Willmer_2006}.
Values of evolution parameters $P$ and $Q$ as well as the faint end slope $\alpha$ for red galaxies are further tuned to match the LSST forecasting years 1 and 10 redshift distributions of source sample galaxy distributions using the LF-based sample-PZ modelling described in Sec. \ref{subsec:convolution_method} and are indicated with a $\dag$ symbol.}
\begin{tabular}{cccccc}
\hline
Galaxy type & $\phi_0^* \; [(h / \mathrm{Mpc})^3]$ & $P$ & $M_0^* (0.1)$ & $Q$ & $\alpha$ \\ \hline
red galaxies & $0.0111$ & $-1.95\dag$ & $-20.34$ & $1.8$\dag & $-0.65\dag$ \\
all galaxies & $0.0094$ & $1.8$ & $-20.70$ & $0.7$ & $-1.23$ \\ \hline
\end{tabular}\label{tab:GAMADEEP2_lf_params}
\end{table*}

\subsection{Redshift distribution}
\label{subsec:convolution_method}

Arguably, the most fundamental quantity that is necessary for any cosmological and/or astrophysical analysis is an accurate and precise measurement of distances to objects.
Cosmological surveys in general employ an approach in which the proxy for the distance (the redshift) is inferred on imaging or spectroscopic data using spectral energy distribution (SED) models.	
These two approaches are conceptually different methods to study the features of the SED.
We can observe large parts of the sky using photometric observations in broad optical filters.
However, the long exposure times required to measure spectra accurately for faint objects can be prohibitively long \citep{2015APh....63...81N}.
As a result, even with modern multi-object spectrographs, we can only expect to obtain spectra for a few percent of the sources which will be imaged by current and next-generation deep and wide photometric surveys \citep{2019NatAs...3..212S}. 
Since the filters in broad-band optical surveys like Rubin are broad and sparse, a lot of the SED information is simply not captured. As a result information on redshift, or distance, is lost by utilizing these lower resolution photometric measurements, as compared with a spectroscopic measurement (up to 100 times, \citealt{2019NatAs...3..212S}).

Redshift estimates that have been obtained using photometric data, referred to as \textit{photometric redshifts}, are commonly employed in cosmological surveys which rely on large data samples, particularly for weak lensing sources. 
An SED-based photometric redshift (PZ hereafter) estimation method is essentially a mapping between per-band fluxes (hence colors) and the galaxy redshifts.
There are several procedures to estimate photometric redshifts, such as template fitting or data-driven methods where machine learning (ML) training algorithms are utilized (for a comprehensive review, consult \citealt{2019NatAs...3..212S}).
Accuracy and precision is demanded by Stage IV surveys, as mentioned above, and understanding the photometric redshift uncertainties are of key importance for achieving these demands.
Uncertainties from photometric redshift estimates propagate into the analysis and ultimately alter the cosmological parameter constraints and our understanding of the underlying cosmological model.

Despite the tremendous community efforts, photometric redshifts are still one of the major systematics in cosmological analyses.
One of the main reasons is that most techniques perform best when the training set is representative of the whole population, which is increasingly difficult to ensure as we push to fainter imaging surveys.
This can then lead to biases.

An alternative to SED fitting or SED-based machine learning methods are the so-called clustering methods \citep{Matthews_2010, Schulz_2010,  McQuinn_2013, Johnson_2016, Lee_2016} that produce independent redshift information. 
These methods introduce sensitivity to galaxy bias into photometric redshift estimation. 
While this can break certain degeneracies, it introduces a layer of complexity not found in SED-based methods.

Another approach for achieving a redshift distribution estimate was developed by \citet{Sheth_2010} and further extended by \citet{van_Daalen_2018}. 
This method provides a mechanism for deriving a redshift distribution, denoted as $\frac{\d N}{\d z}$, via the appropriate combination of a luminosity function $\Phi(M, z)$ with a redshift-dependent survey volume element. We now introduce this method in more detail.

The central expression is  \citep{van_Daalen_2018}:
\begin{equation}\label{eq:dndz_basic_equation}
     \langle N_{i, \lambda} \rangle^{\mathrm{Poiss}} = \int^{z_{i, \mathrm{max}}}_{z_{i, \mathrm{min}}} \d z \int^{m_{\lambda,\mathrm{max}}}_{m_{\lambda, \mathrm{min}}}  \d m \frac{\d \Phi (M(m, z), z)}{\d z} \frac{\d V_C(z)}{\d  z}
\end{equation}
with $\langle N_{i, \lambda} \rangle^{\mathrm{Poiss}} $ the Poisson mean number of galaxies in redshift bin $i$ and apparent magnitude bin $\lambda$. $z_{i, \mathrm{min}}$ and $z_{i, \mathrm{max}}$ are the lower and upper limits of redshift bin $i$ respectively, and similar for $m_{\lambda, \mathrm{min}}$ and $m_{\lambda, \mathrm{max}}$ with respect to apparent magnitude bin $\lambda$.

\begin{figure}
    \centering
    \includegraphics[width=\columnwidth]
    {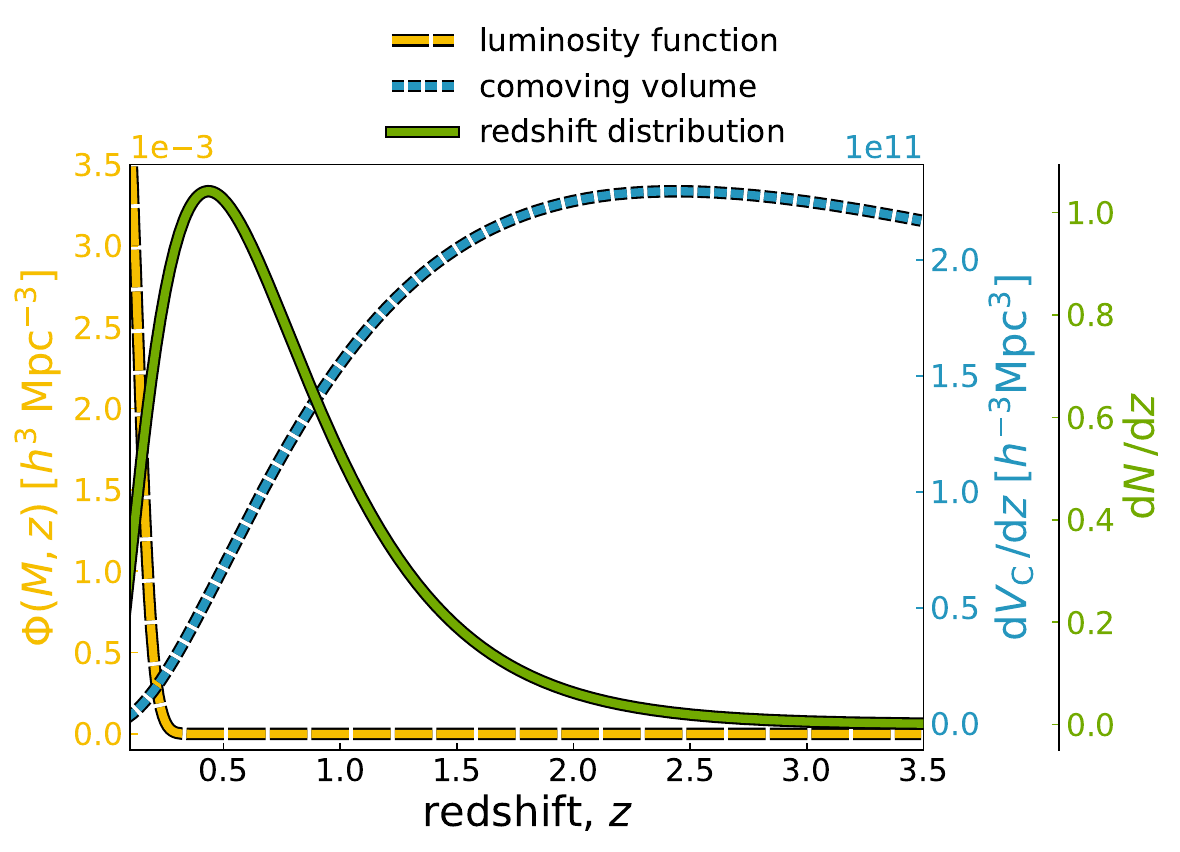}
    \caption{\textit{Illustration of the approach used to model the redshift distribution of source galaxies in this work}.
    The luminosity function as calculated using Eq. \ref{eq:luminosity_function_z_scaling} with fiducial parameters is shown in dashed yellow (evaluated for this illustrative example at $m=17.4$ for the purpose of visualizing the redshift dependence).
    The derivative with respect to redshift of the comoving survey volume, $V_C$ (as can be found using Eq. \ref{eq:comoving_volume}), is shown as a short-dashed teal line.
    Combining these two quantities as in Eq. \ref{eq:dndz_basic_equation} produces the galaxy redshift distribution, represented by the green solid line and normalized here such that it peaks at unity.}
    
    \label{fig:convolution_method_illustration}
\end{figure}

Eq. \ref{eq:dndz_basic_equation} is constructed by integrating the number density (as related to the redshift-dependent luminosity function) over the comoving volume element of the survey on the appropriate magnitude and redshift range (for details, see \citealt{van_Daalen_2018}). The comoving survey volume element, $d\,V_C$, enters Eq. \ref{eq:dndz_basic_equation} via the $\frac{\d V_C(z)}{\d  z}$ term. It is defined through the comoving distance $D_\mathrm{C}(z)$ and the area of the sky covered by the survey $A_{\mathrm{sky}}$:
\begin{align} \label{eq:comoving_volume}
    d\,V_C & =A_{\mathrm{sky}}\, D_{\mathrm{C}}^2 (z) d D_\mathrm{C} \notag \\
    & =4 \pi f_{\mathrm{sky}}   \frac{D_{\mathrm{C}}^2 (z)c}{H_0 \sqrt{\Omega_{m, 0}(1 + z)^3 + \Omega_{\Lambda, 0}}} \d z \notag \\
    &=\frac{dV_C(z)}{dz} dz
\end{align}
with $f_{\mathrm{sky}}$ being the fraction of the sky, $c$ the speed of light in vacuum, $H_0$ the Hubble parameter, and $\Omega_{m,0}$ and $\Omega_{\Lambda,0}$ matter and dark energy density fractions (evaluated at present time), respectively.
A flat $\Lambda$CDM framework is assumed. 

An illustration of the application of Eq. \ref{eq:dndz_basic_equation} can be seen in Fig. \ref{fig:convolution_method_illustration}. 
Fundamentally, the number of galaxies within a given luminosity interval diminishes with increasing redshift, as outlined in Eq. \ref{eq:luminosity_function_z_scaling} — this is shown by the long-dashed yellow line (for a fixed apparent magnitude for illustrative purposes).
Conversely, the derivative of the survey volume with respect to $z$, which is depicted as the short-dashed teal solid line in Fig. \ref{fig:convolution_method_illustration}, grows with redshift.
The result of these two behaviors yields, via Eq. \ref{eq:dndz_basic_equation}, the redshift distribution of galaxies, represented by the green solid line.

The method described in this subsection enables the direct modelling of the redshift distribution of a magnitude-limited galaxy survey via two physically meaningful ingredients: the luminosity function and the survey volume.
Making use of this modelling allows us to connect the modelling of the redshift distribution to other luminosity-dependent effects like intrinsic alignment. 
It also naturally allows for an extension to incorporate directly within the modelling other survey selection functions beyond a volume selection.
\subsection{Intrinsic alignment of galaxies}
\label{subsec:ia_theory}
The framework proposed in this work can be applied to any intrinsic alignment model which incorporates luminosity dependence. Here, we choose the non-linear linear alignment (NLA) model \citep{Bridle_2007}. We adopt a redshift- and luminosity-dependent parametrization for the NLA IA amplitude $A$, following \citet{joachimi2011constraints} and \citet{Krause_2015}: 
\begin{align}
A (L, z) & = A_0 \frac{C_1 \rho_{m,0}}{D (z)}  \left(  \frac{L}{L_0} \right)^{\beta} \left( \frac{1 + z}{1 + z_0^{\mathrm{low}}} \right)^{\eta_z^{\mathrm{low}}} \nonumber \\
& \times \left[ \Theta (z_0^{\mathrm{high}} - z) + \Theta (z - z_0^{\mathrm{high}}) \left( \frac{1 + z}{1 + z_0^{\mathrm{high}}} \right)^{\eta_z^{\mathrm{high}}} \right].
\label{eq:ia_amplitude_model_lf_z}
\end{align}
$A_0$, $\beta$, $\eta_z^{\mathrm{low}}$, and $\eta_z^{\mathrm{high}}$ are free parameters of the model to be constrained, respectively capturing information about the amplitude of the IA signal, its scaling with luminosity, and its scaling with redshift in lower and higher $z$ regimes. $z_0^{\mathrm{low}}$ is a fixed pivot redshift of the low-redshift power law, while $z_0^{\mathrm{high}}$ defines the redshift at which the higher-redshift behavior (defined by $\eta_z^{\mathrm{high}}$) becomes important; similarly $L_0$ is a fixed pivot luminosity. The fiducial values for these parameters used in our forecasting analysis are provided in Table \ref{tab:analysis_fiducial_values} (with $L_0$ given in terms of equivalent absolute magnitude, $M_{\rm piv}$).
$C_1$ is a normalization constant (fixed to $C_1 \rho_{\mathrm{cr}} \approx 0.0134$ via SuperCOSMOS observations \citep{hirata2004intrinsic}, with $\rho_\mathrm{cr}$ the critical density), and $\rho_{\mathrm{m},0}$ is the matter density at $z = 0$ (today). $D(z)$ is the linear growth factor, and $\Theta$ is a Heaviside step function.

We follow \citet{Krause_2015} to obtain the IA amplitude as a function of redshift and survey limiting magnitude, by integrating over the luminosity function for the population of red galaxies:
\begin{equation}
    A (m_\mathrm{lim}, z) = \biggl \langle A (L, z) \biggr \rangle_{{\phi_{\mathrm{
red}}}}{ f_{\rm red}(z)}.
\label{eq:A_mlim_z}
\end{equation}
Here, $m_\mathrm{lim}$ is the survey limiting magnitude and $f_{\rm red}(z)$ is the fraction of red galaxies as a function of redshift. Explicitly, the angular brackets in Eq. \ref{eq:A_mlim_z} are evaluated as:

\begin{align}
 \biggl \langle A (L, &z) \biggr \rangle_{\phi_{\mathrm{
red}}} {= A (L_0, z) \left\langle \left( \frac{L}{L_0} \right)^{\beta} \right\rangle_{\phi_{\mathrm{red}}} \left( \frac{1 + z}{1 + z_0^{\mathrm{low}}} \right)^{\eta_z^{\mathrm{low}}}} \nonumber \\ &{ \times \left[ \Theta (z_0^{\mathrm{high}} - z) + \Theta (z - z_0^{\mathrm{high}}) \left( \frac{1 + z}{1 + z_0^{\mathrm{high}}} \right)^{\eta_z^{\mathrm{high}}} \right]} \nonumber \\ &{ =A (L_0, z) \frac{\int^{\infty}_{L(m_{\mathrm{lim}}, z)} \d L \left( \frac{L}{L_0} \right)^\beta \phi_\mathrm{red} (L, z)}{\int^{\infty}_{L(m_{\mathrm{lim}}, z)} \d L \phi_\mathrm{red} (L, z)}\left( \frac{1 + z}{1 + z_0^{\mathrm{low}}} \right)^{\eta_z^{\mathrm{low}}}} \nonumber \\ &{ \times \left[ \Theta (z_0^{\mathrm{high}} - z) + \Theta (z - z_0^{\mathrm{high}}) \left( \frac{1 + z}{1 + z_0^{\mathrm{high}}} \right)^{\eta_z^{\mathrm{high}}} \right]}.
 \label{eq:ia_amplitude_lumfunc}
\end{align}
 Finally, the fraction of red galaxies, is obtained via:
\begin{equation}
    f_\mathrm{red}{(z)} = \frac{\int^{\infty}_{L(m_{\mathrm{lim}}, z)} \d L  \phi_\mathrm{red} (L, z)}{\int^{\infty}_{L(m_{\mathrm{lim}}, z)} \d L \phi_\mathrm{all} (L, z)}
    \label{eq:red_frac_def}
\end{equation}
with $\phi_\mathrm{red}$ and $\phi_\mathrm{all}$ the luminosity functions of red and all galaxies in the sample.

An important emphasis is that the IA amplitude is modelled as luminosity-dependent. 
Thus, when computing the cosmic shear spectrum from theory, it is necessary to assume a luminosity function parametrization.
Previous forecasting work using this parametrization has fixed the luminosity function parameters to best-fit values measured from external data sets \citep{Krause_2015, srd_lsst_desc}. 
However, it is not clear that this is an appropriate analysis choice; we will investigate this as part of the development of our analysis framework below in Sec. \ref{sec:forecasting_setup}. 
\subsection{Cosmic shear}
\label{subsec:cosmic_shear}

The main observable in this work is the cosmic shear power spectrum.
The observed cosmic shear power spectrum $C _{\epsilon \epsilon}^{ij} (\ell)$ can be modelled as the sum of the gravitational lensing part GG, cross-term GI between lensing and intrinsic alignment, and the pure intrinsic alignment contribution II:\begin{equation}\label{eq:2D_angular_spectrum}
    C_{\epsilon \epsilon}^{ij} (\ell) = C_{\mathrm{GG}}^{ij} (\ell) + C_{\mathrm{GI}}^{ij} (\ell) + C_{\mathrm{IG}}^{ij} (\ell) + C_{\mathrm{II}}^{ij} (\ell) .
\end{equation}
Each contribution is described in spherical harmonic space where $\ell$ is the angular frequency (a 2D wave vector perpendicular to the line of sight) while indices $(i, j)$ represent tomographic redshift bins where the shear correlations are computed (for a review of key intrinsic alignment quantities and notation, see \citet{ia_guide_lamman}).

In this work, we will take our fiducial IA model to be the non-linear alignment (NLA) model \citep{Bridle_2007, Krause_2015}.
We stress that although it is likely that a more complex model of IA may be required for LSST analysis (\textit{e.g}. the TATT model from \citealt{tatt_Blazek_2019}), the NLA model provides a suitable means of incorporating luminosity-dependent IA modelling and is appropriate in the context of a first exploration of our joint modelling methodology.
For the case of the non-linear alignment model of intrinsic alignments, each term in the observed cosmic shear power spectrum is given by:
\begin{align}
    C_{\mathrm{GG}}^{ij} (\ell) = \int_{0}^{\chi_{\mathrm{H}}} \d \chi \frac{q^i(\chi) q^j(\chi)}{f_k^2(\chi)} P_{\delta \delta} \left( \frac{\ell}{f_k(\chi)}, \chi \right) , \\
    C_{\mathrm{GI}}^{ij} (\ell) = \int_{0}^{\chi_{\mathrm{H}}} \d \chi \frac{q^i(\chi) p^j(\chi)}{f_k^2(\chi)} P_{\delta \mathrm{I}} \left( \frac{\ell}{f_k(\chi)}, \chi \right) , \\
    C_{\mathrm{IG}}^{ij} (\ell) = \int_{0}^{\chi_{\mathrm{H}}} \d \chi \frac{p^i(\chi) q^j(\chi)}{f_k^2(\chi)} P_{\mathrm{I} \delta} \left( \frac{\ell}{f_k(\chi)}, \chi \right) , \\
    C_{\mathrm{II}}^{ij} (\ell) = \int_{0}^{\chi_{\mathrm{H}}} \d \chi \frac{p^i(\chi) p^j(\chi)}{f_k^2(\chi)} P_{\mathrm{II}} \left( \frac{\ell}{f_k(\chi)}, \chi \right) .
\end{align}
where $\chi_\mathrm{H}$ is the comoving distance to the horizon, $f_k (\chi)$ is the radial function which in a flat Universe (as we consider in this work) equals the comoving distance $\chi$, and we assume the Limber approximation \citep{limber_Kitching_2017, limber_LoVerde_2008, limber_Lemos_2017}. $q^i (\chi)$ and $q^j (\chi)$ are the lensing weighting functions in tomographic bins $i$ and $j$, respectively: 
\begin{equation}
       q^{i(j)} (\chi) = \frac{3 H_0^2 \Omega_\mathrm{m}}{2 c^2} \frac{f_K (\chi)}{a (\chi)} \int_{\chi}^{\chi_{\mathrm{H}}} \d \chi' p^{i(j)} (\chi') \frac{f_K (\chi' - \chi)}{f_K (\chi')}
\end{equation}
with $H_0$ the Hubble parameter, $\Omega_\mathrm{m}$ the matter density fraction, $c$ the speed of light in a vacuum, and $a(\chi)$ the scale factor. $p^{i(j)}(\chi'(z))$ are galaxy redshift distributions in tomographic bins $i$ or $j$; $p(\chi'(z))^{i(j)} = p(z)^{i(j)} \frac{\d z}{\d \chi}$.
$P_{\delta \delta}$ is the nonlinear power spectrum of the density field, while $P_{\delta \mathrm{I}}$ and $P_{\mathrm{II}}$ are given in the linear and non-linear alignment model of IA by:
\begin{align}
    P_{\mathrm{II}} (k, z) = A^2 (z) P_{\epsilon \epsilon} (k, z), \\
    P_{\mathrm{GI}} (k, z) = - A (z) P_{\delta \epsilon} (k, z)
\end{align}
where in the linear alignment model, $P_{\epsilon \epsilon}$ and $P_{\delta \epsilon}$ are given by the linear matter power spectrum, whilst the non-linear alignment model uses the non-linear matter power spectrum.
Both linear and non-linear matter power spectra are expressed in the Fourier space, $k$.
$A(z)$ is the redshift-dependent intrinsic alignment amplitude that also depends on the galaxy properties and we use the amplitude parametrization presented in Sec. \ref{subsec:ia_theory}.
\subsection{Joint modelling via the luminosity function}
\label{sec:jmas_via_lf}

Given the introduction of the method of modelling the redshift distribution of galaxies via a volume selection and luminosity function in Section \ref{subsec:convolution_method} and of the luminosity-dependent modelling of IA in Section \ref{subsec:cosmic_shear} above, we are in a position to explicitly introduce our proposed joint modelling framework. 

The key point is to take advantage of the physical link of the luminosity function, which is present in the modelling of both the redshift distribution and the intrinsic alignment of galaxies. 
Rather than either 1) fixing the luminosity function parameters to their best-fit values from an external survey in modelling IA or 2) neglecting the explicit luminosity dependence of the IA amplitude, we promote the parameters of the luminosity function to full dimensions of the parameters space in the cosmological inference analysis.
These parameters enter the modelling of both the IA amplitude and of the photometric redshift distribution.
We conjecture that this scenario may allow for the self-calibration of the LF parameters.
It may also enable the breaking of degeneracies between the modelling of IA and of the photometric redshift distribution (as well as other astrophysical systematics were the framework to be extended to include them, like the galaxy bias).
In Section \ref{sec:results}, we will present the results of our initial investigation as to the utility of our proposed method for these and other purposes.

\subsection{Fisher forecasting}
\label{sec:fisher_theory}

We explore the potential of our proposed joint-modelling parameter estimation framework via a Fisher forecasting analysis, chosen for its convenience and efficiency.
In the Fisher forecasting formalism, random variable $\boldsymbol{x}$ (an $N$-dimensional vector) has an associated probability density distribution $L (\boldsymbol{x}; \boldsymbol{\theta})$ that depends on $M$ model parameters $\boldsymbol{\theta} = (\theta_1, \theta_2, ... , \theta_M)$.
The central quantity is the Fisher matrix, defined as:
\begin{equation}\label{eq:fisher_matrix}
    F_{\alpha \beta} \equiv \Big\langle \frac{\partial^2 \mathscr{L}}{\partial \theta_\alpha \partial \theta_\beta} \Big\rangle
\end{equation}
with
\begin{equation}
    \mathscr{L} \equiv - \ln L.
\end{equation}
Assuming that the probability distribution is a multivariate Gaussian, one can derive an analytical formula by examining the derivatives and their expectation values $\boldsymbol{\mu} \equiv \langle \boldsymbol{x} \rangle$ \citep{Tegmark_1997}
\begin{equation}\label{eq:fisher_matrix_derivs}
    F_{\alpha \beta} = \langle \mathscr{L}_{, \alpha \beta} \rangle   = \frac{1}{2} \mathrm{Tr} \left[\boldsymbol{C}^{-1} \frac{\partial \boldsymbol{C}}{\partial \theta_\alpha} \boldsymbol{C}^{-1} \frac{\partial\boldsymbol{C}}{\partial \theta_\beta} \right] + \frac{\partial \boldsymbol{\mu}^t}{\partial \theta_\alpha} \boldsymbol{C}^{-1} \frac{\partial \boldsymbol{\mu}}{\partial \theta_\beta}
\end{equation} 
with $\boldsymbol{C} \equiv \langle \boldsymbol{x x}^t \rangle - \boldsymbol{\mu \mu}^t$ the covariance matrix and all quantities being evaluated at a set of fiducial values of the parameters $\boldsymbol{\theta}$.
In the case in which the covariance matrix does not depend appreciably on the parameters $\boldsymbol{\theta}$, the above reduces to:
\begin{equation}\label{eq:fisher_matrix_fisk}
    F_{\alpha \beta} = \frac{\partial \boldsymbol{\mu}^t}{\partial \theta_\alpha} \boldsymbol{C}^{-1} \frac{\partial \boldsymbol{\mu}}{\partial \theta_\beta}
\end{equation} 
which is the formulation of the Fisher matrix we use in this work.
According to the multivariate Cramér-Rao inequality \citep{rao1992information}, the inverse of the Fisher matrix is the lower bound on the covariance matrix of the cosmological parameters (under the assumption that our analysis set-up allows for an unbiased estimate of the parameters from data): 
\begin{equation}
 \text{cov}(\hat{\boldsymbol{\theta}}) \geq \boldsymbol{F}^{-1} \, .   
\end{equation}
Here $\hat{\boldsymbol{\theta}}$ denotes the vector of estimated parameter values and $\boldsymbol{F}^{-1}$ denotes the inverse of the Fisher information matrix. 

We note that in this work we incorporate prior information in our Fisher forecasts (with the parameters of these Gaussian priors given below in Table \ref{tab:analysis_fiducial_values}). We follow the methodology of e.g. \citet{bassett2011fisher} in assuming such prior information is independent from the information offered in the present analysis, such that we can simply add Fisher matrices representing this prior information to our Fisher matrix before inverting to obtain estimated (lower bounds on the) parameter covariance.

In our study, we acknowledge the inherent instabilities in Fisher forecasting analyses due to the reliance on numerical derivatives, as highlighted in \citet{bhandari2021fisher}.
In order to ensure the numerical stability, we employ the \textit{"stem method"}, as detailed in \citet{Camera_2016}, which has been effectively applied in Euclid forecasts \citep{EuclidForecast2020, Bonici_2023}. 
A detailed explanation of the implementation of this method into our framework is given in Sec. \ref{sec:forecasting_setup} on Fisher forecasting infrastructure.
\section{Forecasting set-up}\label{sec:forecasting_setup}

We now summarize the analysis choices and major framework building blocks which underpin our results.
The forecasts we conduct are for the cosmic shear probe for LSST Year 1 and Year 10 (Y1 and Y10 hereafter).
For a comprehensive overview of the relevant parameters and their values, please refer to Table \ref{tab:analysis_fiducial_values}.

\begin{table*}
\centering
\caption{Modelling choices used in this analysis. 
For parameters which are varied in the analysis, the values given are fiducial.
If not explicitly indicated, the central values are kept the same for both forecasting years.
Values in parentheses are the $1\sigma$ width of the Gaussian prior imposed on this parameter in the analysis unless otherwise stated.
No value in parentheses indicates a flat prior. 
Luminosity function parameters for `all' galaxies are the same as indicated in Table \ref{tab:GAMADEEP2_lf_params}.}
\label{tab:analysis_fiducial_values}
\begin{tabular}{ccccccc}
\hline
\multicolumn{7}{c}{\textbf{Cosmological parameters}} \\ \hline
$\Omega_m$($\sigma_{\Omega_m}$) & $\sigma_8$($\sigma_{\sigma_8}$) & $n_s$($\sigma_{n_s}$) & $w_0$($\sigma_{w_0}$) & $w_a$($\sigma_{w_a}$) & $\Omega_b$($\sigma_{\Omega_b}$) & $h$($\sigma_{h}$) \\ \hline
$0.3156(0.15)$ & $0.831( 0.2)$ & $0.9645(0.1)$ & $-1.0( 0.5)$ & $0.0(1.3)$ & $0.0491685(0.005)$ & $0.6727(0.125)$ \\ \hline
\multicolumn{7}{c}{\textbf{Luminosity function parameters (red galaxies)}} \\ \hline
forecast year & $\phi_0^* \: [h^3 \mathrm{Mpc^{-3}}]$ & $P$ & $M_0^*$ & $Q$ & $\alpha$ & $z_0$ \\ \hline
1 & $0.0111$ & $-2.20$ & $-20.34$ & $1.20$ & $-0.98$ & $0.10$ \\
10 & $0.0111$ & $-1.85$ & $-20.34$ & $1.80$ & $-1.10$ & $0.10$ \\ \hline
\multicolumn{7}{c}{\textbf{Intrinsic alignment parameters}} \\ \hline
$A_0$ & $\beta$ & $\eta_z^\mathrm{low}$ & $\eta_z^\mathrm{high}$ & $z_0^\mathrm{low}$ & $z_0^\mathrm{high}$ & $M_\mathrm{piv}$ \\ \hline
5.92(2.5) & $1.10(1.0)$ & $-0.47(1.5)$ & $0.00(0.5)$ & $0.30$ & $0.75$ & $-22.00$ \\ \hline
\hline
\end{tabular}
\end{table*}

\textbf{Cosmological model.}
We base our forecasts on spatially flat $\Lambda$CDM model cosmologies,  utilizing the following set of cosmological parameters: the total present matter density parameter $\Omega_m$ (with $\Omega_m = \Omega_b + \Omega_c$); $\Omega_b$ the baryon fraction; $\sigma_8$ the root mean square of the amplitude of matter density perturbations at an $8 \: h^{-1} \: \mathrm{Mpc^{-1}}$ scale; $h$ the reduced Hubble constant ($H_0 = h \: 100 \: \mathrm{km \: s^{-1} \: Mpc^{-1}}$); $n_s$ the primordial scalar perturbation spectral index.
We adopt the Chevallier-Polarski-Linder (CPL) model for parametrization of dark energy's evolution with redshift $z$ \citep{CPL_w0wa, linder_expansionhistory_2003PhRvL..90i1301L}, represented by the equation of state:
\begin{equation}
    w (z) = w_0 + w_a \frac{z}{1 + z}.
\end{equation}
The cosmological parameters we varied in our analysis, along with their respective fiducial values, are presented in Table \ref{tab:analysis_fiducial_values}.

In calculating the cosmic shear power spectrum, we use the Core Cosmology Library \citep[\pkg{pyCCL v2.8.0}]{Chisari_2019}, with the linear matter power spectrum calculated via \pkg{CAMB} \citep{Lewis_2000}. 
Baryonic corrections are not applied.

\textbf{Luminosity function model.}
We allow a single set of luminosity function parameters to describe the `red' luminosity function of Eqs. \ref{eq:ia_amplitude_lumfunc} and \ref{eq:red_frac_def} (used in modelling IA) and the total luminosity function of Eq. \ref{eq:dndz_basic_equation} (used in modelling the redshift distribution). We take this set of parameters as free parameters to be constrained within a cosmic shear analysis. For the parameters of the `all' luminosity function of  Eqs. \ref{eq:ia_amplitude_lumfunc} and \ref{eq:red_frac_def} (used in modelling IA), we use the `all' parameters of GAMA directly (given in Table \ref{tab:analysis_fiducial_values}) and do not vary these in our analysis. This set-up is appropriate for our objective of a simplified, proof-of-concept analysis in which the IA amplitude and redshift distribution modelling share a set of luminosity function parameters, while the red fraction (and hence the fractional impact of IA on cosmic shear) is kept to a physically expected level for LSST. A more complete and correct treatment would incorporate additionally a selection function for color as a function of luminosity and redshift, with additional free parameters to be constrained; we leave the explicit incorporation of such selection functions for future work.
 
Given this set-up, we require fiducial values of free parameters of our luminosity function which achieve both the redshift distributions and IA amplitude which agree with LSST expectations as defined in \citet{srd_lsst_desc}. 
We follow \citet{Krause_2015} in using parameter values as fit to the GAMA survey as a starting point (representing as they do a viable physical luminosity function, albeit from a survey with a very different selection).
Given this considerable difference in survey selection (with GAMA having a much brighter limiting magnitude than that expected for LSST Y1 or Y10), it is not surprising that directly implementing these GAMA luminosity function parameters does not result in redshift distributions or IA amplitudes which match those expected for LSST (illustrated in Fig. \ref{fig:nz_w_gama_lfs}). 
Given this, we iteratively adjust the luminosity function parameters  to achieve a fiducial luminosity function which, within our modelling framework, produces good agreement with the redshift distribution and redshift-dependent IA amplitude expected for LSST. The refined values of these parameters, tailored to meet the forecasting scenarios, are detailed in Table \ref{tab:analysis_fiducial_values}.

\begin{figure*}
\includegraphics[width=0.8\textwidth]
{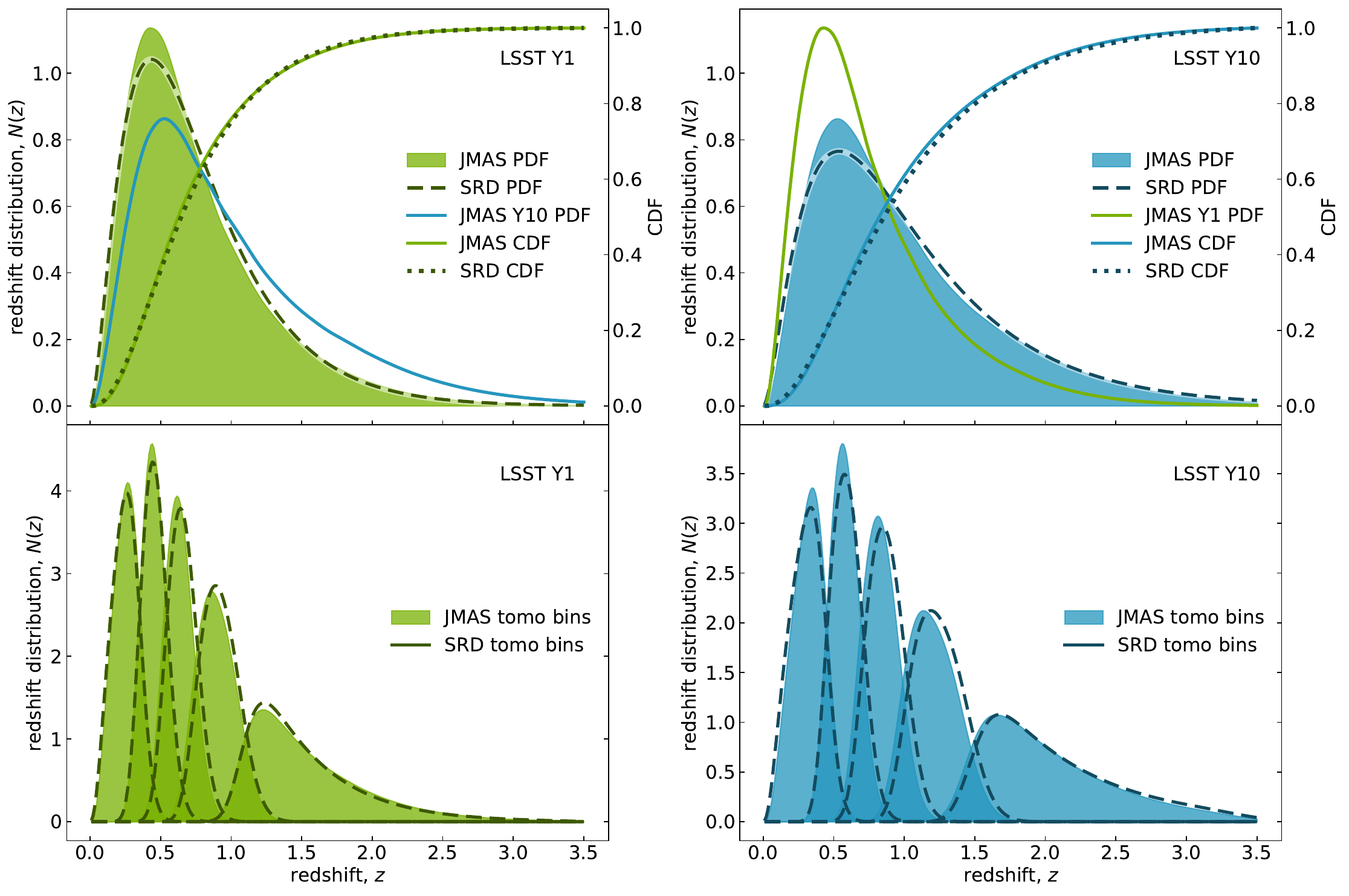}
\caption{\textit{Source sample redshift distributions and tomographic bins used in this work.}
\textit{Top panels}: Normalized source galaxies redshift distributions for LSST Y1 (left) and Y10 (right).
Redshift distributions from \citet{srd_lsst_desc} (in dashed lines, labeled as "SRD", "Science Requirement Document"), against which we compare our fiducial distributions computed using the modelling method described in Sec. \ref{subsec:convolution_method}, are represented in solid lines in both plots.
Corresponding redshift distributions used in our analyses (labeled as "JMAS","Joint Modelling of Astrophysics Systematics") are represented in green (Y1) and blue (Y10).
The success of our implementation is corroborated by comparing the respective cumulative distribution functions (CDFs), indicated by dotted lines for CDFs from \citet{srd_lsst_desc} and solid green and blue lines for the method of this work, for Y1 and Y10, respectively. 
We also highlight the difference in the redshift distributions between Y1 and Y10 scenarios by including Y10 distribution (blue solid line) in the Y1 panel (top left) and Y1 distribution (green solid line) in the Y10 panel (top right).
This difference plays a major role in our approach when interpreting the Fisher forecasting results, as highlighted in Sec. \ref{subsec:jmas_results}
\textit{Bottom panel}: SRD (dashed lines) and JMAS tomographic bins for LSST Y1 and Y10 (green and blue shaded regions, respectively).}
\label{fig:srd_jmas_dndz_and_binning}
\end{figure*}

\textbf{Redshift distribution and tomographic binning.}
We allow for the possibility that the galaxies in our cosmic shear analyses (source galaxies, `sources' hereafter) span a redshift range of $0.01 \le z \le 3.51$.
In both forecast years, the redshift range step size is 0.01.
The total redshift distribution of source galaxies as given in the LSST Science Requirements Document (SRD) is parametrized as:
\begin{equation}
    \frac{\d N}{\d z} \propto {\left( \frac{z}{z_0} \right)}^2 \exp{\left[-\left(\frac{z}{z_0}  \right)^{\alpha}\right]}
\end{equation}
with  $(z_0, \alpha) = (0.13, 0.78)$ for Y1 and $(z_0, \alpha) =(0.11, 0.68)$ for Y10 \citep{srd_lsst_desc}.
In our joint-modelling framework, we instead model this redshift distribution via the method by \citet{van_Daalen_2018}, described in Sec. \ref{subsec:convolution_method}.
We thus need to ensure that our fiducial luminosity function parameters reproduce a realistic underlying, unbinned source galaxy redshift distribution, considering the survey \textit{i}-band limiting magnitudes of 24.1 and 25.3 for forecasting years 1 and 10, respectively.
Our procedure for doing so is as follows: we start from a luminosity function in a Schechter form.
We initially take a set of LF parameters for $r$-band red galaxies from the GAMA survey \citep{Loveday_2011}. 
From there, we use the insights gained from exploring the LF parameter space (explained in detail in Section \ref{sec:lf_params_impact_dndz}) to find a set of fiducial LF parameters which produce redshift distributions which match the LSST Y1 and Y10 distributions as given in \citet{srd_lsst_desc}. 
Specifically, we match the mean, mode, and median redshifts to the greatest extent possible. 
Apart from that, we inspect the corresponding cumulative distribution functions (CDFs) to ensure that the overall distribution shape aligns well with our expectations and the parameters of the study.
To achieve this, we adjusted the density parameter $P$, rate of change of the pivot magnitude (luminosity) $Q$, as well as the faint end slope $\alpha$ from their initial GAMA survey values. 
The precise values and adjustments for these parameters can be found in Table \ref{tab:analysis_fiducial_values}.
The redshift distributions we generated are consistent with the LSST SRD ones to a high degree, as can be seen in the upper panels of Fig. \ref{fig:srd_jmas_dndz_and_binning}.

For both Y1 and Y10 forecasts, we then follow \citet{srd_lsst_desc} to create five tomographic bins and to incorporate effects on the binning from photometric redshift uncertainty. 
We first segment the overall true redshift distribution into five equipopulated tomographic bins by applying top-hat window functions constructed on this basis, $T_i(z)$, for each bin $i$. 
 We then take the simplifying assumption that the joint probability distribution of the true redshift with the photometric redshift in each tomographic bin $i$ conforms to a Gaussian distribution at each redshift point (see, for example, \citealt{MaHuHuterer_2006}):
\begin{equation}
p^i (z_{\mathrm{PZ}} | z) = \frac{1}{\sqrt{2 \pi} {\sigma_z^i}} \exp \left[ - \frac{(z - z_{\mathrm{PZ}} - z_{\mathrm{bias}}^i)^2}{2 (\sigma_z^i)^2} \right].  
\end{equation}
with $z_{\mathrm{bias}}^i$ representing for redshift bin $i$ a shift in the mean of the Gaussian and away from $z = z_{PZ}$ and $\sigma_z^i$ representing the possibly-z-dependent variance term.
Given this, we have the redshift distribution $\frac{dN}{dz}^i$ of tomographic bin $i$ to be:
\begin{equation}
    \frac{dN}{dz}^i = \frac{\int dz \frac{dN}{dz} T^i(z) p^i(z_{\mathrm{PZ}}| z)}{\int d z_{\mathrm{PZ}} \int dz \frac{dN}{dz} T^i(z) p^i(z_{\mathrm{PZ}| z)}}
\end{equation}

Each tomographic bin is then described in principle by distinct uncertainty parameters. 
Fiducially, bias on the redshift $z_{\mathrm{bias}}^i$ is set to zero in each bin $i$ while scatter in the LSST case is modelled for all bins as a linear function of redshift
\begin{equation}
    \sigma_z^i = 0.05 (1 + z)
\end{equation}
for both forecast years. 
Although it is conventional to vary one or both of these sets of parameters in cosmological analysis, we fix them here. This is because we instead allow to vary the luminosity parameters which produce to underlying redshift distribution.
The resulting redshift distributions are visualized in the bottom panels of Fig. \ref{fig:srd_jmas_dndz_and_binning}, using green color for Y1 and blue for Y10.
We also over-plot the LSST SRD bins (in dashed lines) for completeness and easy comparison.

\begin{figure}
\centering
\includegraphics[width=\columnwidth]{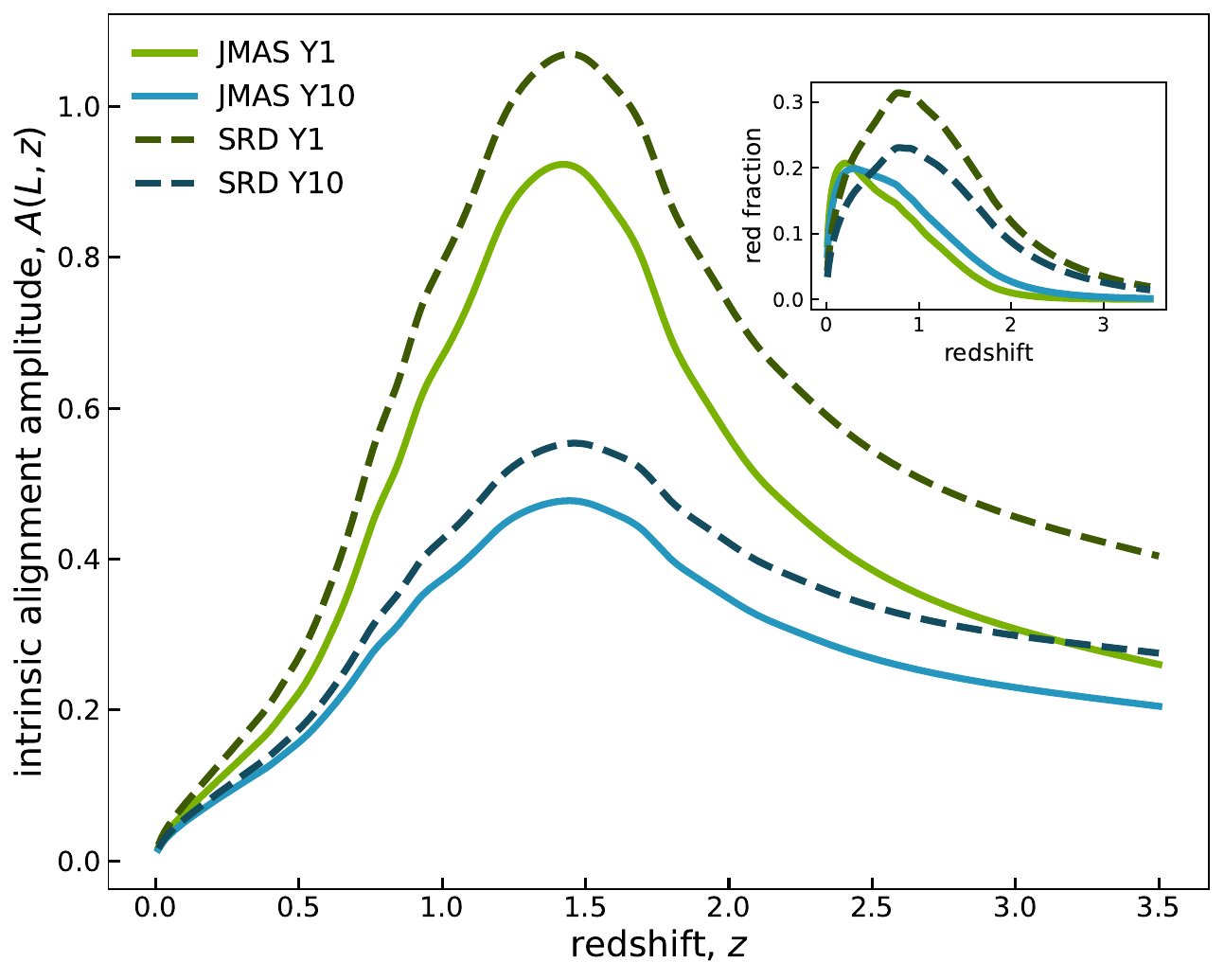}
 \caption{
 \textit{The redshift evolution of the fiducial IA amplitude $A (L, z)$}.
The IA amplitude is calculated using the luminosity function with the fiducial parameter values provided in Table \ref{tab:analysis_fiducial_values}.
The Y1 case from this work (`JMAS Y1') is represented by a green solid line, contrasting with the JMAS Y10 scenario's blue solid line. 
Dashed lines illustrate the SRD cases for comparison. 
Notably, the JMAS Y1 IA amplitude exceeds that of Y10, primarily because the latter captures a larger fraction of fainter galaxies, which typically exhibit weaker intrinsic alignments.
 Both JMAS and SRD IA amplitudes display similar distribution peaks, with minor discrepancies due to differences in redshift distribution and luminosity function parameters.
The IA amplitude calculation factors in the fraction of red galaxies, leading to divergence  between the JMAS and SRD scenarios.
An inset visualizes this: Y1 forecast (green) and Y10 (blue) for JMAS are in solid lines, while SRD results are dashed. 
The red galaxy fraction in the JMAS case peaks earlier, and overall amplitudes are lower, influenced by the adjusted luminosity function parameters affecting galaxy density $P$, rate of change of the pivot luminosity $Q$, and the faint end slope $\alpha$.
}
\label{fig:ia_amplitude}
\end{figure}

\textbf{Intrinsic alignment model.}
For the intrinsic alignments, our model incorporated a luminosity and redshift-dependent amplitude, as detailed in Eq. \ref{eq:ia_amplitude_model_lf_z}. The fiducial intrinsic alignment parameters --- $A_0$, $\beta$, $\eta_z^{\mathrm{low}}$, and $\eta_z^{\mathrm{high}}$ --- employed in our analysis are the fiducial values of \citet{srd_lsst_desc} and can be found in Table \ref{tab:analysis_fiducial_values}.
The fiducial redshift evolution of the IA amplitudes as well as the corresponding red fraction can be viewed in Fig. \ref{fig:ia_amplitude}: the solid green line represents the amplitude for Y1 while Y10 is represented in blue.
The intrinsic alignment amplitude in both survey scenarios exhibits a sharp increase up to a redshift of approximately 1.5, beyond which it demonstrates a more gradual decline.
Notably, the overall amplitude observed in the Y1 sample surpasses that projected for the year 10 forecasting scenario.
For completeness and easier comparison, we included the redshift evolution of the IA signal for the corresponding SRD cases in Fig. \ref{fig:ia_amplitude}.

In our analysis, we consciously retained the same color and evolutionary corrections as those utilized in obtaining the SRD results \citep{cosmolike_Krause_2017, srd_lsst_desc}.
This decision was driven by the objective to minimize alterations to the established parameters, ensuring that the focus remains on evaluating the method's effectiveness and validity.
This approach aligns with the intent of our investigation, which is to rigorously test the capabilities and accuracy of the method within a consistent and controlled framework.

\textbf{Data vector.}
Our data vector consists of auto- and cross-correlation angular power spectra $C_\ell$ of cosmic shear.
Considering there are 5 bins (Fig. \ref{fig:srd_jmas_dndz_and_binning}) in both Y1 and Y10, there are 15 angular power spectra in total.
For the multipole range, we follow \citet{srd_lsst_desc} in considering only $\ell \le 3000$. 
Our particular choice of $\ell$ bins was constructed to match that of \citet{srd_lsst_desc}, to enable us to take advantage of data products from that work, particularly covariance matrices. Specifically: 20 $\ell$ bins are defined to be logarithmically spaced within the range $20 \le \ell \le 15,000$, prior to then performing the above cuts to restrict to $\ell \leq 3000$.
We do not model baryons in our analyses.
As with the redshift distribution, we employ a validation step and compare the jointly modelled data vectors to those of the standard analysis (fiducial data vectors from \citealt{srdtarball}). 
In this process, we examine the ratio of the power spectra ${C^{ij}_{\mathrm{JMAS}} (\ell)}$ and $ {C^{ij}_{\mathrm{SRD}} (\ell)}$ for each correlation, performing this analysis separately for both forecasting years 1 and 10. 
Specifically, we calculate the relative difference between the JMAS and SRD data vectors for each correlation (for both forecasting scenarios) to assess the relative deviations.
Our findings are reassuring; the deviations do not exceed $15\%$ in any case, indicating that it is possible to achieve a high degree of consistency between data vectors modelled in our framework and those calculated within a more standard modelling set-up.

\textbf{Covariance matrix.}
To ensure consistency, we have utilized the covariance matrices from the SRD study, as detailed in \citet{srd_lsst_desc}.
These matrices, along with other SRD resources, are publicly available and can be found in \citet{srdtarball}.
For an in-depth understanding of the construction and characteristics of these covariance matrices, readers are referred to the comprehensive descriptions in \citet{srd_lsst_desc} and the original work by \citet{cosmolike_Krause_2017}.

\begin{figure}
\centering
\includegraphics[width=\columnwidth]{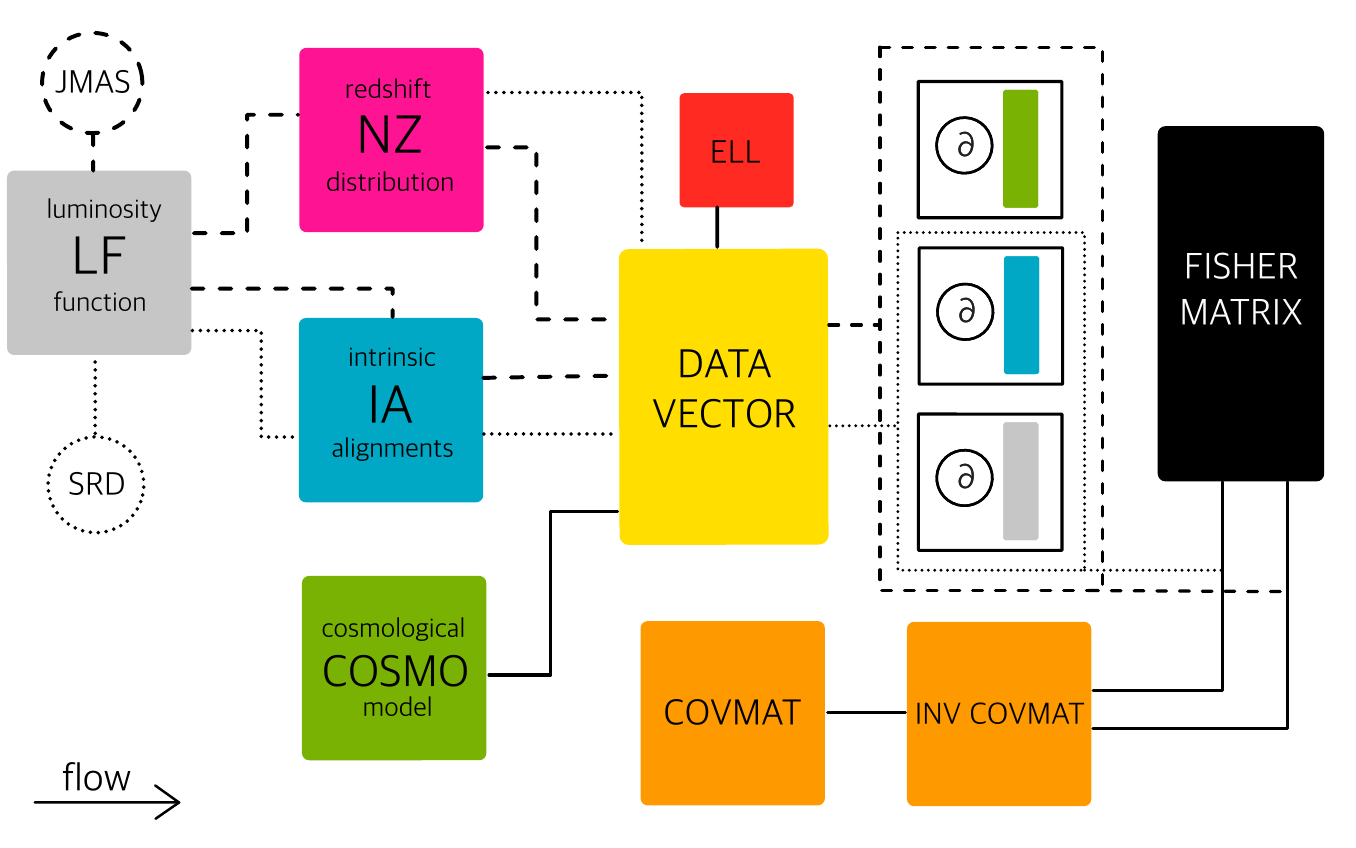}
 \caption{\textit{FISK Architecture and Workflow.} This diagram illustrates the main components of the Fisher forecasting pipeline utilized in our study. 
 The cosmological model (COSMO) and astrophysical systematic models (redshift distribution NZ, intrinsic alignment IA, and luminosity function LF) construct the primary observable -- the data vector (angular power spectra), using \pkg{pyCCL}. 
 Subsequently, the derivatives of the data vectors are calculated using the \textit{stem} method.
 These are then combined with the inverse of the covariance matrix to generate the Fisher matrix.
 The paths taken in the standard forecast (SRD) are depicted with dotted lines, while the joint modeling framework (JMAS) is shown with dashed lines.
 Solid lines represent the common pathway for both SRD and JMAS forecasts.
 The pipeline is designed for flexibility, allowing the user to specify the desired forecast (SRD or JMAS), the forecasting year (1 or 10), or to alter any of the model parameters (COSMO, IA, NZ, LF).}
\label{fig:fisk_algo}
\end{figure}

\textbf{Fisher forecasting infrastructure.} To perform Fisher analysis, we have developed a specialized code, FISK (Fisher Inference for Systematics in Kosmology).
This code is designed as an object-oriented pipeline, incorporating a range of structural elements that align with the key components discussed in earlier sections of this document. 
At its core, FISK consists of various classes corresponding to each model type, including NZ (Redshift Distribution), IA (Intrinsic Alignment), LF (luminosity function), and a cosmological model.
These classes contribute to the construction of the primary observable, the cosmic shear data vector using \pkg{pyCCL}.
Based on user input, the pipeline can execute a standard analysis (SRD) or follow the path of joint modelling (JMAS). 
Additionally, as we will demonstrate, there is a third option (SRD+LF) that maintains the SRD framework while incorporating derivatives over LF parameters.
The calculation of derivatives of the data vector with respect to model parameters is handled by a dedicated `Derivative' class. 
Lastly, the calculated derivatives are integrated with the inverse of the covariance matrix, resulting in the construction of the Fisher matrix.
The FISK algorithm is illustrated in Fig. \ref{fig:fisk_algo}.
The dashed line in Fig. \ref{fig:fisk_algo} shows the route of the JMAS framework while the SRD standard analysis is depicted with a dotted line.
The solid line represents the mode common for both forecasting modes (SRD and JMAS).
The user can also specify the forecasting year (1 or 10) or, if desired, change the value of the fiducial parameters for all the models of the study.

As noted in Sec. \ref{sec:fisher_theory}, we encountered numerical instabilities when calculating derivatives using the five-point stencil or methods available in the \pkg{numdifftools} \citep{numdifftools} Python library.
To address the limitations of traditional finite difference techniques, we implemented a \textit{stem} method, as detailed in \citet{Camera_2016}. 
Following their approach, we calculate the data vector $C^{ij}(\ell)$ at 15 points surrounding each model fiducial parameter $\theta_\alpha$: $\delta \theta_\alpha= 0, \pm0.625\%, \pm1.25\%, \pm1.875\%, \pm2.5\%, \pm3.75\%, \pm5\%, \pm10\%$. 
For parameters with a fiducial value of zero (\textit{e.g.}, $w_a$), we use the percentage changes directly, without scaling by $\theta_\alpha$, ensuring the method remains robust even in such cases.
Considering the anticipation of linear results, we strive to maintain the fit’s deviation from the actual values within our targeted precision of 1\%. 
Should this level of precision not be met, edge data points are excluded, and the fitting process iterated until the desired precision is attained. Upon fulfilling the precision criteria, we define the derivative as the slope of the linear fit.
This method has proven highly effective and stable against variations in redshift range resolution or other factors influencing the angular power spectra.
Our thorough testing of this derivative calculation approach (see Figs. \ref{fig:corner_srd_stem_test_z_res} and \ref{fig:corner_srd_stem_test_z_res_pert}) demonstrates its robustness and superiority over the finite difference method for the analysis case we present here.
Additionally, we have conducted extensive validations of our Fisher matrices by comparing them with the Fisher matrices released in the LSST DESC Science Requirements Document \citep{srdtarball}.

In addition to its core functionality, FISK is thoughtfully designed to allow for the generation of intermediate results at any stage of the pipeline.
This feature is particularly beneficial for validation purposes, as it enables thorough examination and verification of each step in the process. 
Furthermore, the code is designed to be easily extendable, modifiable, and upgradable.
This flexible architecture ensures that FISK can readily adapt to evolving investigation, accommodate new models or methodologies, and integrate advancements in future analyses (for example, including the modelling of the galaxy bias).
\section{Results}
\label{sec:results}

\subsection{Understanding the impact of the luminosity function parameters on the redshift distribution}
\label{sec:lf_params_impact_dndz}

We start by examining how changes in the parameters of the luminosity function affect the redshift distribution using the method described in Section \ref{subsec:convolution_method}. 
Understanding these effects will help us interpret later results and create realistic LSST-like redshift distributions as outlined in Section \ref{sec:forecasting_setup}.

\begin{figure*}
\includegraphics[width=\textwidth]{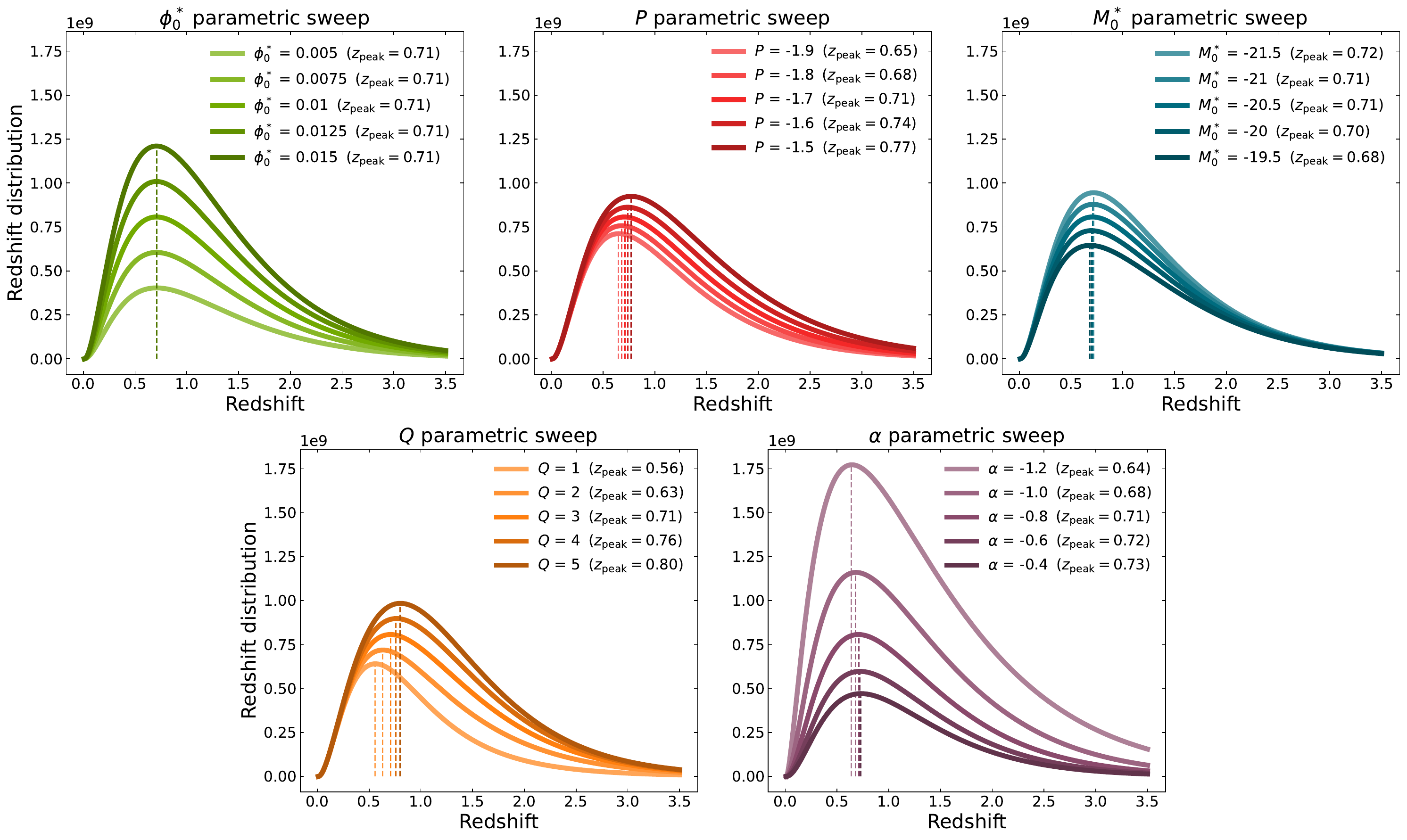}
    \caption{\textit{The effect of luminosity function parameters on a magnitude-limited redshift distribution.}
    This figure demonstrates the influence of each Schechter luminosity function parameter on the redshift distribution. 
    Each LF parameter was varied while keeping others fixed.
    The normalization factor ($\phi_{0}^*$, in green solid lines) primarily adjusts the amplitude, the density evolution parameter ($P$, top middle panel) affects both amplitude and peak shift towards higher redshifts, and the characteristic magnitude ($M_{0}^*$ in blue) influences the amplitude and peak position, favoring lower redshifts for higher values. 
    The rate of change $Q$ (in orange-yellow) emphasizes brighter galaxies at higher redshifts, while the faint end slope ($\alpha$, in purple) lowers amplitude and shifts the peak to higher redshifts. 
    We intentionally excluded color and evolutionary corrections from these distributions to maintain simplicity and focus solely on the impact of the LF parameters.
    This approach allows for a clear demonstration of the fundamental effects, avoiding the additional complexity and fluctuations introduced by these corrections (which are present due to the redshift dependency of the evolutionary corrections).
    Our aim was to present a straightforward, showcase scenario, emphasizing the core aspects of the LF parameters without the influence of more intricate, redshift-dependent factors.}
    \label{fig:lf_param_sweep_dndz}
\end{figure*}

In our analysis, we employ the LF-based sample-PZ modelling methodology within the LSST survey framework.
Our objective was to understand the influences of individual Schechter luminosity function parameters on the redshift distribution, a critical aspect for understanding results of our Fisher forecast further on.
Each parameter was varied in isolation, with others held constant at their central values, enabling a focused investigation into the unique effects of each factor.
The findings are shown in Fig. \ref{fig:lf_param_sweep_dndz}.

\begin{enumerate}
    \item The normalization factor, $\phi_0^*$, which denotes the number density of galaxies at a benchmark luminosity and redshift, predominantly modulates the overall amplitude of the redshift distribution.
    This parameter acts as a scaling factor, and its variation is indicative of changes in the overall galaxy population density over cosmic time. 
    The parametric sweep over the normalization parameter is indicated in green solid lines in Fig. \ref{fig:lf_param_sweep_dndz}.
    An increase in $\phi_0^*$ suggests a denser universe with a higher overall number of galaxies, amplifying the amplitude of the distribution.
    \item In the context of the luminosity function's influence on the redshift distribution of galaxies, the parameter $P$, which encapsulates the concept of galaxy density evolution with redshift, exerts a multifaceted impact.
    Specifically, $P$ modulates both the amplitude and the peak positioning of the redshift distribution.
    An increase in the $P$ value indicates a heightened density of galaxies at higher redshifts, as illustrated by the red solid lines in Figure \ref{fig:lf_param_sweep_dndz}.
    This rise in density not only boosts the total count of galaxies within the observed distribution but also leads to a notable shift in the distribution's peak towards higher redshifts.
    Such a shift, facilitated by a higher $P$ value, describes a more compact arrangement of galaxies, leading to lower lacunarity within our fixed volume element.
    \item In the analysis exploring the interplay between the luminosity function and the redshift distribution of galaxies, a key observation centers around the "knee" of the luminosity function, characterized by the parameter $M_0^*$. 
    We have systematically increased $M_0^*$ from -22 to -18 in increments of 1, as shown in blue solid lines in Fig. \ref{fig:lf_param_sweep_dndz}.
    Notably, as the characteristic magnitude increased (indicating a shift towards dimmer characteristic luminosities), there is a marked decrease in the amplitude of the galaxy redshift distribution.
    Concurrently, a shift of the peak of the redshift distribution towards lower redshifts with larger $M_0^*$ values is present.
    This suggests that galaxies around and below the "knee" (dimmer galaxies) are more prevalent or detectable at lower redshifts, significantly influencing the shape and peak of the redshift distribution.
    \item In the context of galaxy evolution and distribution, a higher value of the parameter $Q$ in the luminosity function indicates a scenario where intrinsically brighter galaxies are more prevalent at higher redshifts. The role of this parameter is to modulate the evolution of the characteristic magnitude, $M^*$, across redshifts, following Eq. \ref{eq:lf_params}.
    This is reflected in the redshift distribution of galaxies (orange-yellow solid lines in Fig. \ref{fig:lf_param_sweep_dndz}), where a higher $Q$ value results in a distribution that is skewed towards these more luminous galaxies at elevated redshifts.
    This trend is observable as an enhancement in both the peak and amplitude of the redshift distribution, indicating a higher concentration of bright (and hence for our apparent-magnitude-limited survey, observable) galaxies in the ancient universe.
    \item When modelling the redshift distribution of galaxies based on the luminosity function, an increase in the (fiducially negative) $\alpha$ value (indicating a flatter faint-end slope and hence a smaller number of low-luminosity galaxies) is observed to cause a decrease in the amplitude of the redshift distribution (distributions in purple in Fig. \ref{fig:lf_param_sweep_dndz}).
    Lower-luminosity galaxies make up a substantial portion of the overall galaxy number density across all redshifts. Although there is some corresponding increase in galaxies at luminosities just above the pivot, this effect is largely washed out by the dominance of the exponential term at higher luminosities. Since we have fixed the normalisation parameter $\phi_0^*$ for this exercise, it then follows that reducing the number of intrinsically faint galaxies will reduce the overall amplitude of the redshift distribution.
    Additionally, there is a noted shift in the peak of the redshift distribution towards higher redshifts as $\alpha$ increases towards zero.
    This shift could be attributed to a minor but still present increased relative representation of higher-luminosity galaxies (still in the faint-end i.e. below pivot luminosity), recalling that at higher redshifts we are preferentially going to observe intrinsically brighter galaxies. 
\end{enumerate}

In summary, these parameters collectively characterize the galaxy population, each contributing to the shape of the redshift distribution of galaxies over cosmic time. 
Furthermore, it is evident from the mathematical expressions (Eq. \ref{eq:lf_params}) that there exists a degree of degeneracy among these parameters. 
Such degeneracies are critical in understanding the limitations and constraints of our model and interpretations.
The exact nature and directions of these degeneracies will be explored in the forthcoming discussion on our Fisher analysis results. 
\subsection{Understanding the impact of the luminosity function parameters on the cosmic shear angular power spectra}
\label{sec:lf_params_impact_datavector}

\begin{figure*}
\includegraphics[width=\textwidth]
{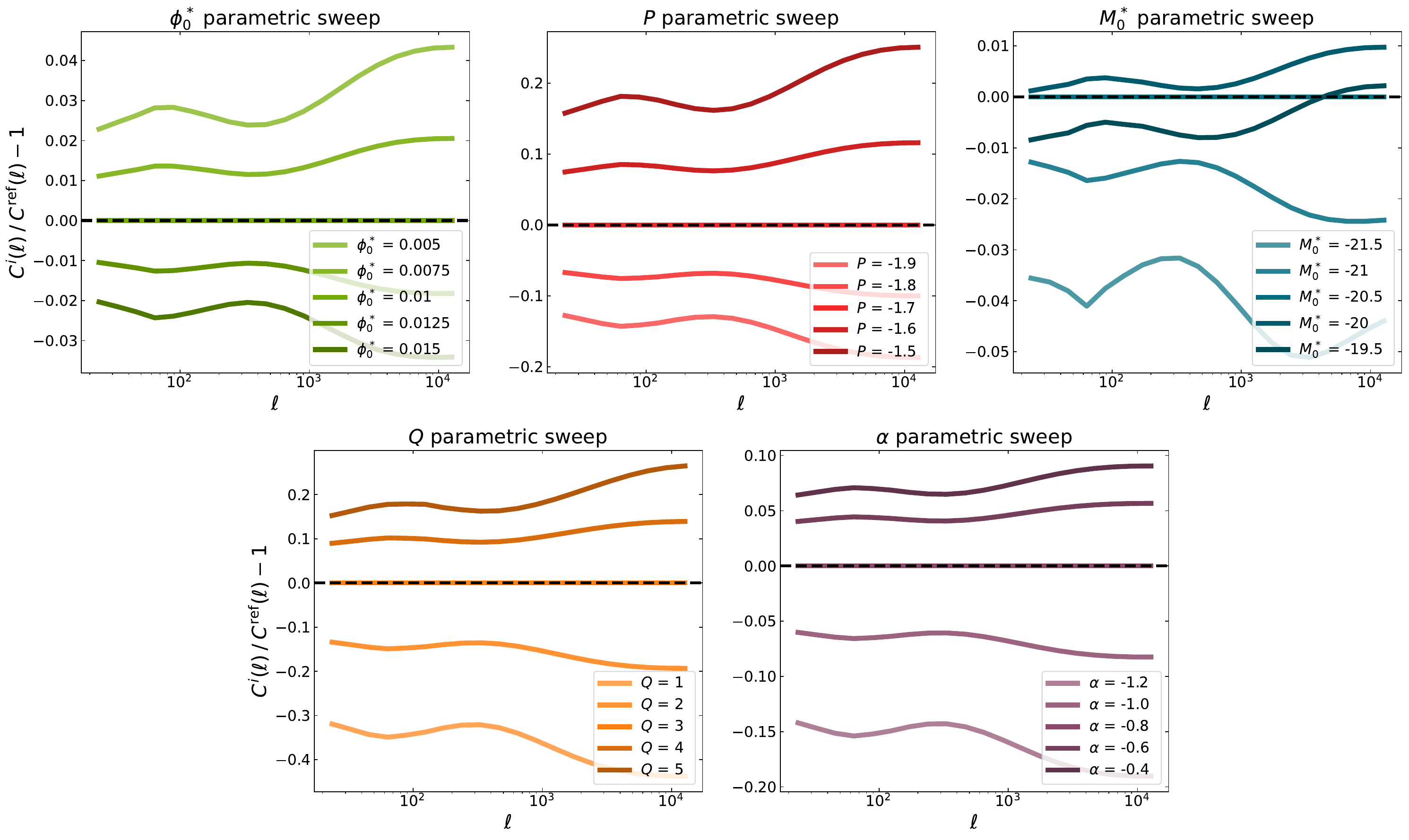}
\caption{{\textit{Impact of Luminosity Function Parameters on Cosmic Shear Angular Power Spectra.} This figure illustrates the relative difference in cosmic shear angular power spectra $C_\ell$, as a function of various Schechter luminosity function parameters.
LF parameters are varied over five values, with the central value serving as the reference for relative comparison.
In each plot, only the first bin correlations are displayed (bin 1- bin 1 correlation).
The normalization factor $\phi_0^*$ is shown in green solid lines, highlighting that an increase leads to a negative relative difference ratio.
The density evolution parameter $P$ is depicted in red solid lines, where an increase results in a higher density of galaxies at higher redshifts, thus augmenting the shear signal.
The characteristic magnitude $M_0^*$ is in blue, and the rate of change $Q$ in orange-yellow, both showing a similar positive trend upon increase.
Conversely, an increase in the faint-end slope $\alpha$ (in purple) is observed to also contribute to the shear signal, evident from the deviation pattern across different scales. 
The adjustment of each luminosity function parameter, though executed in equal incremental steps, does not uniformly translate to proportional changes in the amplitude of $C_\ell$.
In some instances it is clear that as we deviate further from the central value, either by increasing or decreasing a parameter, the impact on the amplitude changes disproportionately.
This non-linear response highlights the complex, non-proportional relationship between the variations in luminosity function parameters and their consequent effects on the cosmic shear strength.}}
\label{fig:lf_param_sweep_datavector}
\end{figure*}

Following the methodology of the redshift distribution parametric sweep from the joint modeling method (Sec. \ref{sec:lf_params_impact_dndz}), we extended our analysis to explore the influence of the Schechter luminosity function parameters on our modelled data vector, the cosmic shear angular power spectra $C_\ell$.
This approach entailed computing the jointly modelled angular power spectra for (bin 1 - bin 1) combination for the varied LF parameters and representing the results as the relative difference ratio compared to the reference power spectra. 
Note that because we display the autospectrum for the lowest tomographic z-bin, we expect the IA contribution to be much more significant than in the case of higher $z$ or cross-spectra across separated tomographic bins. 
We choose this deliberately to be able to see the impact of IA as well as of the redshift distribution on cosmic shear, where otherwise the latter would heavily dominate.
For each LF parameter, we explored a range of five distinct values.
The central value among these five was designated as the reference for comparison purposes.
The analysis revealed distinct trends and deviations as a function of multipole $\ell$, contingent upon the variation of specific parameters (while keeping other parameters fixed).
Results are presented in Fig. \ref{fig:lf_param_sweep_datavector}.

  \begin{enumerate}
        \item Increasing the normalisation factor $\phi_0^*$, resulted in the relative difference ratio consistently being below zero (left top panel in Fig. \ref{fig:lf_param_sweep_datavector}).
        Physically, a higher $\phi_0^*$ implies an increased number density of galaxies with respect to luminosity.
        We attribute this effect to the impact of the LF parameter on the IA's contribution to the spectrum.
        As a normalisation factor, the effect of $\phi_0^*$ will ultimately be normalised out of the lensing contribution.
        However, due to the way it is incorporated into the IA amplitude in our model (specifically, the fact that the red fraction, Eq. \ref{eq:red_frac_def} contains a factor of $\phi_0^*$ for the red galaxy selection), this normalisation does not occur.
        So, an increase in $\phi_0^*$ increases the fraction of red galaxies and hence the IA contribution, and thus causes a relative decrease in the overall spectrum.
        \item An increase in the density evolution parameter $P$ leads to an augmented relative difference ratio, as shown in the top middle panel in Fig. \ref{fig:lf_param_sweep_datavector}. 
        This result can be interpreted as being primarily due to the lensing contribution to the spectrum, as increasing $P$ shifts the mean redshift of the tomographic bin up and therefore increases the amplitude of the lensing signal. 
        \item For both the characteristic magnitude $M_0^*$ (in blue in Fig. \ref{fig:lf_param_sweep_datavector}) and the rate of change $Q$ (in orange-yellow in Fig. \ref{fig:lf_param_sweep_datavector}), an increase results in a positive relative difference. 
        This trend implies that shifts in the luminosity function towards brighter galaxies (lower $M_0^*$) or an evolution towards brighter galaxies at higher redshifts (higher $Q$) enhance the gravitational lensing effect, thereby increasing the cosmic shear signal. 
        Decreasing these parameters resulted in a negative relative difference, suggesting a diminished shear signal due to a shift towards fainter galaxies or a less pronounced luminosity evolution.
        \item An increase in (fiducially negative) $\alpha$, the faint-end slope, also increases the shear signal, as can be seen in the bottom right panel in Fig. \ref{fig:lf_param_sweep_datavector}.
        This indicates that a shallower faint-end slope, corresponding to a relatively higher number of high-luminosity galaxies, ultimately contributes to the overall cosmic shear signal (and increases its amplitude). 
    \end{enumerate}
\subsection{Impact of varying luminosity function parameters in IA modelling only}
\label{subsec:results_ia_lf_only}

Before considering a full joint modelling of IA and source photometric redshift distributions via the luminosity function, we consider also the impact of simply marginalizing over the luminosity function parameters of the IA modelling an LSST cosmic shear analysis.
In the case of the LSST Science Requirements Document (SRD) analysis of \citet{srd_lsst_desc} (which we use extensively in this work as a comparison point and an example of a standard analysis set-up), a Schechter luminosity function with fixed parameter values from the GAMA and DEEP2 surveys is used.
The idea of including luminosity function parameters as nuisance IA modelling parameters was explored in \citet{Krause_2015} for an LSST-like survey; here we present the results of doing so in our particular set-up to build intuition prior to introducing the luminosity function jointly within the redshift distribution and IA modelling.

For this, we generated Fisher forecasts that correspond to the scenarios we are interested in at this instance: an analysis which fixes LF parameters in IA modelling (denoted as SRD) and one where they are varied (denoted as SRD+LF).
Each of the two scenarios are explored in terms of the LSST forecasting years 1 and 10, which makes up four analyses in total.
In other words, the forecasting setup is exactly the same for the SRD and SRD+LF case (for corresponding years) and the only difference is that in the SRD+LF scenario, we vary the LF parameters.

\begin{figure}
\centering
\includegraphics[width=\columnwidth]{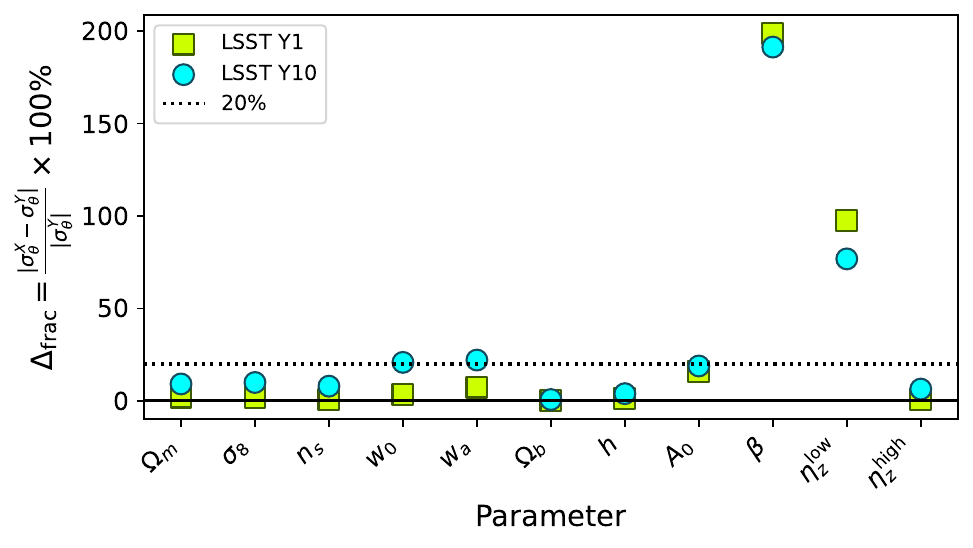}
 \caption{\textit{Fractional Differences between SRD and SRD+LF Scenarios Across Key Parameters.}  
This figure illustrates the impact of including luminosity function parameters in intrinsic alignment (IA) modeling for an LSST cosmic shear analysis, quantified through fractional differences in marginalized $1\sigma$ errors.
Each symbol represents the fractional difference for a specific parameter: neon green squares for Year 1 (Y1) forecasts and neon teal circles for Year 10 (Y10) forecasts.
The plot encompasses seven cosmological parameters and four IA parameters, showcasing their sensitivities to the inclusion of luminosity function parameters in different forecasting years.
Notably, while some parameters like $\Omega_b$ and $h$ show negligible impact across both scenarios, significant effects are observed on $w_0$ and $w_a$ parameters, especially for the Y10 scenario.
For the IA parameters, the fractional differences for $\beta$ and $\eta_z^{\mathrm{high}}$ highlight a substantial sensitivity to the inclusion of LF parameters, with $\beta$ and $\eta_z^{\mathrm{low}}$ showing a decrease in fractional differences between Y1 and Y10.
}
\label{fig:srdlf_stats}
\end{figure}

Since we aim to understand the importance of including luminosity function parameters in the forecast, we analyze the resulting Fisher matrices to extract the $1-\sigma$ errors on parameters.
To quantitatively assess the impact of this inclusion, we compute the fractional difference between the SRD and SRD+LF scenarios for these errors on the parameters. 
This is defined as:
\begin{equation}
\label{eq:fractional_difference}
    \Delta_{\text{frac}} = \frac{|\sigma_\theta^X - \sigma_\theta^Y|}{|\sigma_\theta^Y|} \times 100\% \;.
\end{equation}
In Eq. \ref{eq:fractional_difference}, $\sigma$ represents a $1\sigma$ error on parameter $\theta$, where $X$ and $Y$ denote the forecasting scenarios, with $Y$ being the benchmark.
This approach allows us to specifically highlight the differences in forecast accuracy attributable to the inclusion of luminosity function parameters.
The fractional percentage differences (henceforth referred to as `differences') for each parameter are presented in Fig. \ref{fig:srdlf_stats}. 
The differences for the Year 1 (Y1) forecast are indicated by green squares, while those for Year 10 (Y10) are shown with teal circles.
To aid in interpreting the plot, a line indicating a 20\% difference is included. 

Figure \ref{fig:srdlf_stats} displays the differences for seven cosmological parameters and four IA parameters across both forecasting years. 
It is important to note that LF parameters are not included in the plot, as the SRD \citep{srd_lsst_desc} scenario does not vary these parameters.
However, in the SRD+LF scenario, these LF parameters are indeed varied, as reflected at the beginning of this section.
Analysis of the plot reveals several key insights.
As expected, the errors on the parameters $\Omega_b$ and $h$ are negligibly affected in both Y1 and Y10 forecasting scenarios.
For parameters like $\Omega_b$, $\sigma_8$, and $n_s$, we observe that the errors remain modest for Y1.
However, for Y10, where tighter constraints are expected, the fractional differences increase substantially. 
This observation is particularly significant for $w_0$ and $w_a$, parameters critical to the LSST's science goals as a Stage IV cosmology survey.
The sensitivity of these parameters to the expanded parameter space is notable.
For Y10, the differences for dark energy equation of state parameters exceed 20\%, impacting the survey's objectives.
Regarding the IA parameters, their sensitivity to the inclusion of LF parameters is evident, and expected considering their modelling involves the luminosity function.
The parameter $\eta_z^{\mathrm{high}}$ shows differences below 10\% across both forecasting years, aligning with our limited understanding of the IA amplitude at higher redshifts. 
For $A_0$, including LF parameters in the forecast appears to lead to about 16\% increase in errors for Y1 and 19\% Y10.
The most pronounced impacts are observed in the scaling factor $\beta$ and $\eta_z^{\mathrm{high}}$.
For the $\beta$ parameter, the error increases by 200\% in both forecasting years. 
For $\eta_z^{\mathrm{high}}$, the increase is approximately 100\% for Year 1 (Y1) and 80\% for Year 10 (Y10). 
Notably, the degree of increase in uncertainty for these two parameters shows a relative improvement in Year 10 compared to Year 1. 
This could be potentially attributed to the increased volume of data available in the later observing stages of the LSST, which offers more overall constraining power.

This preliminary investigation underscores the potential benefit of fully integrating luminosity function parameters into cosmological analyses. 
Jointly modelling systematic effects via a shared luminosity function not only leverages underlying correlations but also enriches our understanding of the interconnections among various parameters.
For completeness, we include the Fisher matrix marginalized contours in the Appendix in Fig. \ref{fig:srd_vs_srd+lf_contours}.
\subsection{Forecasts for Luminosity-Function-Based Joint Modelling of IA and N(z)}\label{subsec:jmas_results}

Having established the qualitative effect of varying the luminosity function parameters on the modellng method for redshift distributions (Sec. \ref{sec:lf_params_impact_dndz}), the data vector (Sec. \ref{sec:lf_params_impact_datavector}), and explored the impact on forecast constraints of varying the LF parameters in the IA modelling only (Sec. \ref{subsec:results_ia_lf_only}), we now address the primary question of this work: \textit{what is the impact of jointly modelling IA and the redshift distribution via the luminosity function on cosmological parameter constraints}? 

\begin{figure*}
    \centering
\includegraphics[width=\textwidth]{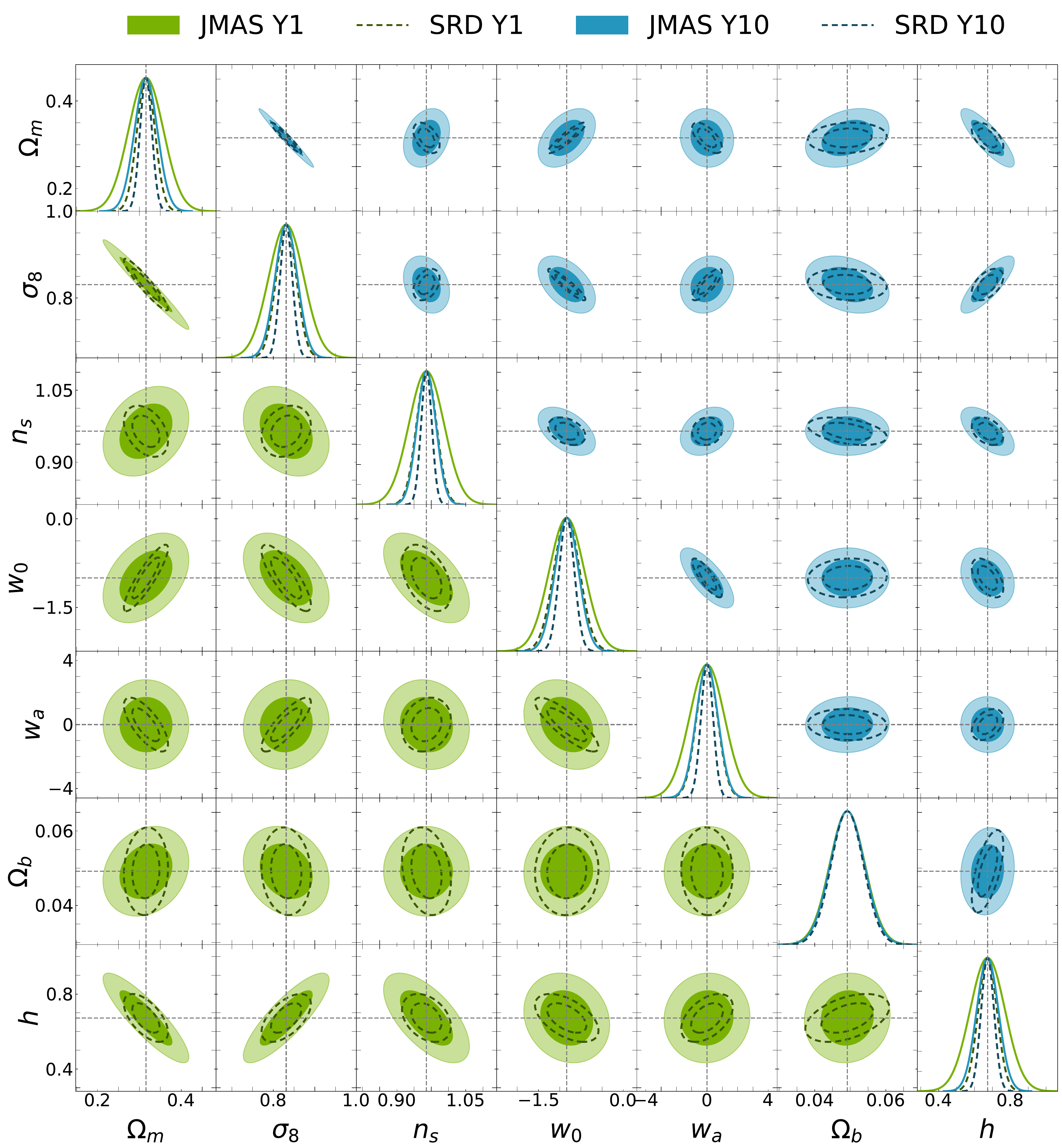}
    \caption{\textit{Fisher matrix marginalized contours for forecasts across different analysis scenarios.} This figure depicts relationships between JMAS (this work) Y1 and SRD Y1 (lower triangle) and JMAS Y10 ad SRD Y10 (upper triangle). 
    The plots contrast the effects of decoupling volume selection and luminosity function on cosmological parameters, with green filled contours representing JMAS Y1, unfilled dark green contours for SRD Y1 (dashed), blue filled contours for JMAS Y10, and unfilled blue dashed contours for SRD Y10. 
    The diagonal elements showcase 1D marginal distributions across all scenarios.}
\label{fig:corner_jmas_vs_srd_cosmo}
\end{figure*}

In this study, we analyze four distinct Fisher forecasting scenarios: two correspond to our integrated framework for modelling astrophysical systematics in the LSST Year 1 and Year 10 forecasts (denoted as JMAS Y1 and JMAS Y10, respectively), and two are based on a standard analysis approach, which treats intrinsic alignment (IA) and redshift distributions independently, as outlined in \citet{srd_lsst_desc}, for LSST Year 1 and Year 10 (referred to as SRD Y1 and SRD Y10, respectively).
The forecasted parameter constraints on cosmological parameters for LSST Year 1 and LSST Year 10 are shown in Fig. \ref{fig:corner_jmas_vs_srd_cosmo}
(the complete parameter space, encompassing the IA and LF parameters, will be thoroughly explored in subsequent sections).
LSST Year 1 posterior constraints are illustrated in the lower segment of the triangle plot in Fig. \ref{fig:corner_jmas_vs_srd_cosmo}, while LSST Year 10 constraints are presented in the upper corner of the figure.
The JMAS Y1 constraints are visualized using light green filled contours, while the SRD Y1 constraints are shown as unfilled dashed contours in dark green. 
For Year 10, the JMAS Y10 scenario is represented with blue filled contours, and the SRD Y10 scenario with unfilled dashed contours in dark blue. 
The diagonal elements present 1D marginal distributions on parameters across all scenarios.

Our key objective is to assess the effect on the constraints of cosmological parameters from 1) decoupling the volume selection from the luminosity function selection in the redshift distribution modeling and 2) considering the latter jointly with intrinsic alignment, especially when the parameter space is expanded. 
The sensitivity of cosmic shear to these modelling details is highlighted by the notable variance in constraining power and the change in degeneracy directions between the JMAS and SRD cases. 
This is particularly apparent in the flipped correlation between $n_s$ and both $\Omega_m$ and $\sigma_8$ parameters see Fig. \ref{fig:corner_jmas_vs_srd_cosmo}.
Such inversions in degeneracy directions can be attributed to the influence of newly introduced nuisance luminosity function parameters, which interact with cosmological parameters.

Despite this, our approach achieves cosmological parameter constraints which are comparable with, if somewhat broader than, cosmological forecasts using more standard methods.
These broader constraints can be viewed to some extent as a more realistic representation of the inherent uncertainties, particularly in relation to the use of a luminosity-function-dependent IA model in both cases.
It is in this context that we delve now into the detailed examination of parameter degeneracies and potential biases.

\subsubsection{Degeneracies:  cosmological and luminosity function parameters}\label{subsec:degeneracies_and_biases_cosmo_vs_lf}

\begin{figure*}
\centering
\begin{subfigure}{0.49\textwidth}
\centering
\includegraphics[height=0.35\textheight]{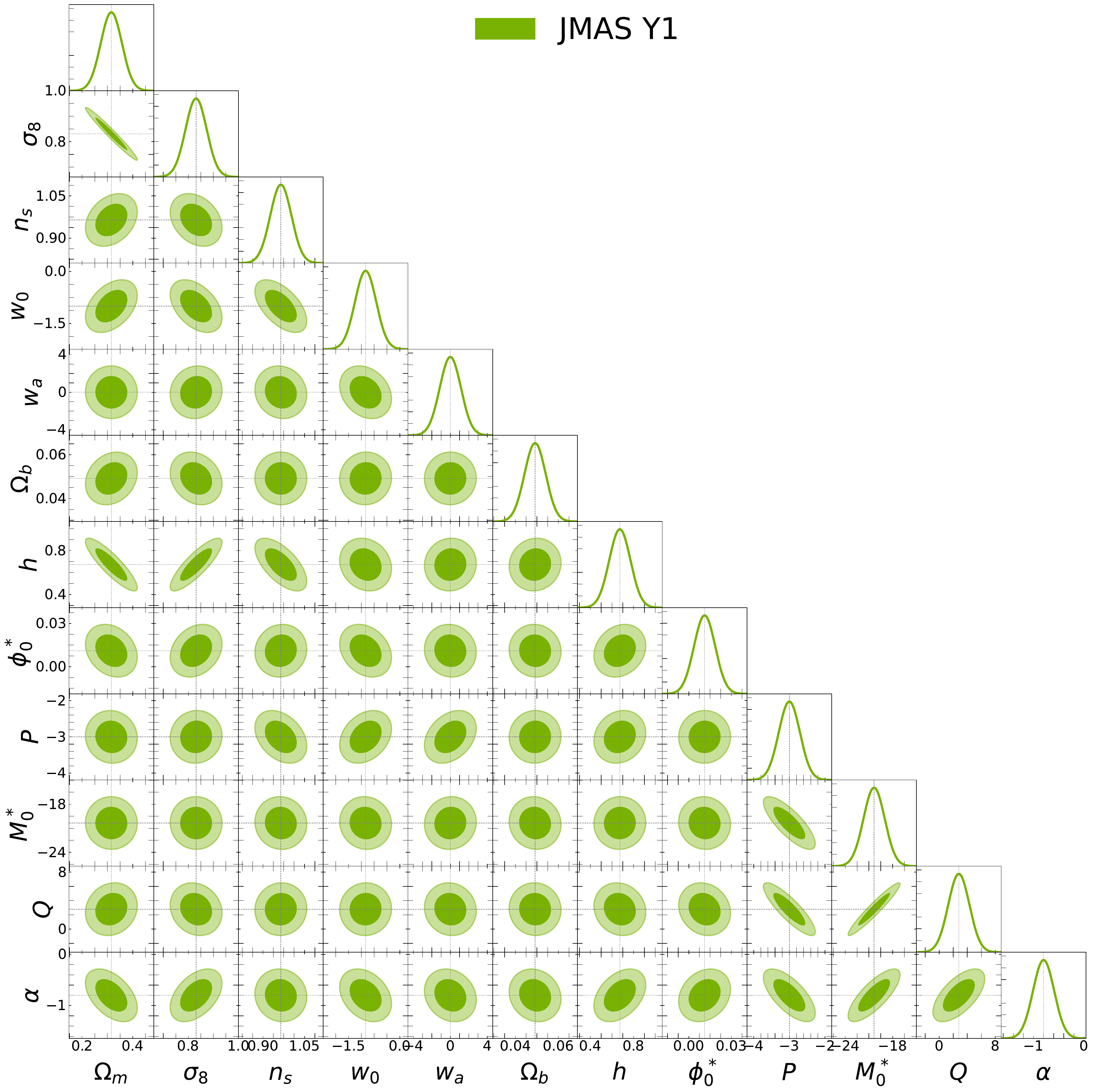}
\caption{Cosmological and LF parameters for LSST Y1.}
\label{fig:cosmolf_y1}
\end{subfigure}
\begin{subfigure}{0.49\textwidth}
\centering
\includegraphics[height=0.35\textheight]{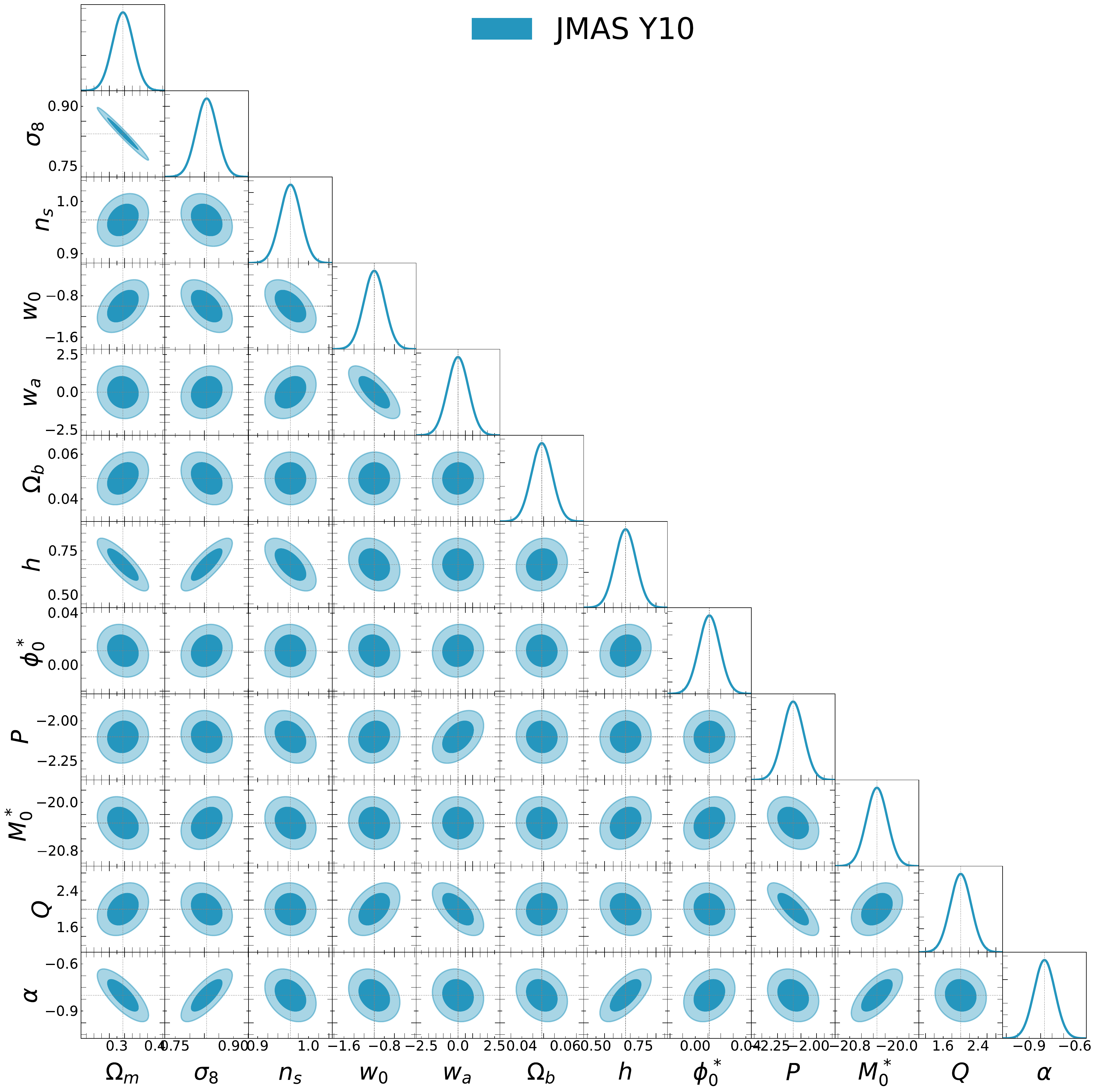}
\caption{Cosmological and LF parameters for LSST Y10.}
\label{fig:cosmolf_y10}
\end{subfigure}
\caption{\textit{Segments of the Fisher matrix marginalized contours for JMAS analysis across forecasting years.}
The left panel illustrates the correlations between cosmological and LF parameters for LSST Year 1 (Y1, green contours), while the right panel shows the same for Year 10 (Y10, blue contours).
This side-by-side arrangement highlights the progression of parameter constraints over time.}
\label{fig:corner_cosmolf_side_by_side}
\end{figure*}

This section explores the degeneracies observed between some key cosmological parameters and the parameters of the luminosity function. 
The corner plots for the complete parameter space, pertaining to the forecast years 1 and 10, are presented in the Appendix, specifically in Figures \ref{fig:corner_jmas_y1_cosmoialf} and \ref{fig:corner_jmas_y10_cosmoialf}.
For a focused analysis, we have extracted and highlighted the segments showcasing the interplay between the cosmological and luminosity function parameters.
These specific segments are presented for comparison in Figure \ref{fig:corner_cosmolf_side_by_side}.

\textbf{Matter density fraction} $\boldsymbol{\Omega_m}$.
We observe a mild anti-correlation between $\Omega_m$ and $\phi_0^*$ particularly in Year 1, where higher source galaxy densities are degenerate with lower overall matter density.
This is interesting as we have seen that the effect of $\phi_0^*$ on the redshift distribution is only to change the normalisation. 
This would thus necessarily result from the impact of $\phi_0^*$ on the IA spectra, where both parameters would impact the overall normalisation.
We see also a clear negative correlation between $\Omega_{\rm m}$ and $\alpha$ in both Year 1 and Year 10, which we can interpret as being in part due to the fact that increasing $\alpha$ (the faint-end slope of the LF) shifts the peak of our redshift distribution to higher $z$.
Increasing the value of $\alpha$ leads to a stronger cosmic shear signal, which can offset the effects of a lower matter density parameter, $\Omega_{\rm m}$.

The increased prominence of the correlation between $\Omega_{\rm m}$ and both $M_0^*$ and $Q$ from Year 1 to Year 10 indicates that, as the survey advances and the statistical power of our data strengthens, the luminosity function of source galaxies plays a more significant role in constraining cosmological parameters.

\textbf{Variance of matter density perturbations} $\boldsymbol{\sigma_8}$.
With respect to $\sigma_8$, we once again see the strongest correlations with $\phi_0^*$ and $\alpha$, with the latter being particularly strong in Year 10. These correlations are both reversed with respect to what we saw between the luminosity function parameters and $\Omega_{\rm m}$, which we attribute to the well-known tight anti-correlation between these two cosmological parameters in cosmic shear constraints. 

\textbf{Dark energy equation of state parameters.}  
We observe degeneracies between both $w_0$ and $w_a$ and several luminosity function parameters, most notably $P$, $Q$, and $\alpha$. The presence of these degeneracies, specifically with luminosity function parameters which shift the mean of the source redshift distribution, suggest that the degree of redshift evolution in the source galaxy luminosity distribution is somewhat degenerate with our time-evolving dark energy model.

\textbf{Spectral index} $\boldsymbol{n_s}$.
In our Fisher analysis with the luminosity function joint modelling approach, we observe consistent but wider constraints between the spectral index $n_s$ and other cosmological parameters.
Notably, there is a reversal in the directions of the ellipses for marginalised 2D constraints of $n_s$ and both $\Omega_m$ and $\sigma_8$.
This reversal is likely influenced by the specific modelling choices and fiducial values of the luminosity function (LF) parameters. 
While $n_s$ remains generally poorly constrained by cosmic shear data, this observation provides valuable insights into the impact of LF modelling on the analysis, highlighting the intricate interplay between LF modelling and cosmological parameter estimation.

\textbf{Dimensionless Hubble parameter} $\boldsymbol{h}$.
Although $h$ is not a parameter traditionally well-constrained by cosmic shear measurements, it does of course impact weak lensing observables. 
We note that in our analysis, $h$ shows a strong correlation with $\alpha$ in both forecasting years, which is tighter in Year 10.
This consistent and strengthening relationship indicates that universes with a shallower faint-end slope of the source galaxy luminosity function, characterized by a smaller number of low-luminosity source galaxies, are degenerate with a higher expansion rate today.
This can be interpreted as being due to the fact that a smaller absolute value of $\alpha$ corresponds to a shift in the source galaxy distribution to higher redshifts and thus a stronger cosmic shear signal.
A stronger cosmic shear signal from this effect can then be compensated by a higher expansion rate, which acts to depress the amplitude of large-scale structure. 

The evolution in some cases of these correlations between Y1 and Y10 highlights the necessity of continually updating our models and assumptions with the progression of observational data.
We note that it is important to bear in mind, when interpreting these results, that fiducial luminosity function parameters differ between Y1 and Y10.
This distinction implies luminosity function parameter constraints will showcase disparate degeneracies and magnitudes, given they are not identical random variables.
That is to say: we do not expect the values of the true luminosity function parameters for LSST Y1 to be the same as for Y10. 

\subsubsection{Degeneracies: intrinsic alignment and luminosity function parameters}\label{subsec:degeneracies_and_biases_ia_vs_lf}

\begin{figure*}
\centering
\begin{subfigure}{0.49\textwidth}
\centering
\includegraphics[height=0.35\textheight]{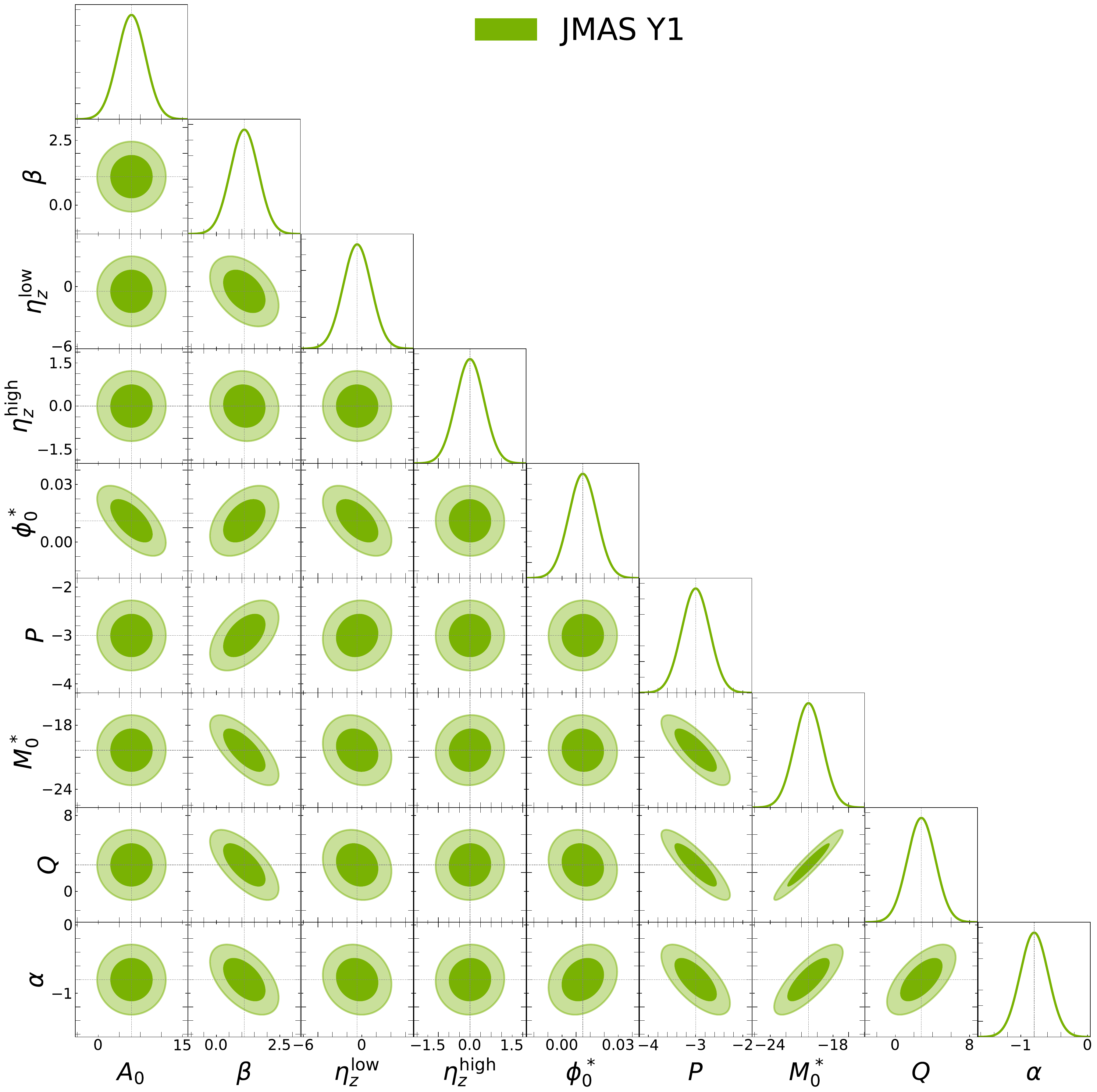}
\caption{IA and LF parameters for LSST Y1.}
\label{fig:ialf_y1}
\end{subfigure}
\hfill
\begin{subfigure}{0.49\textwidth}
\centering
\includegraphics[height=0.35\textheight]{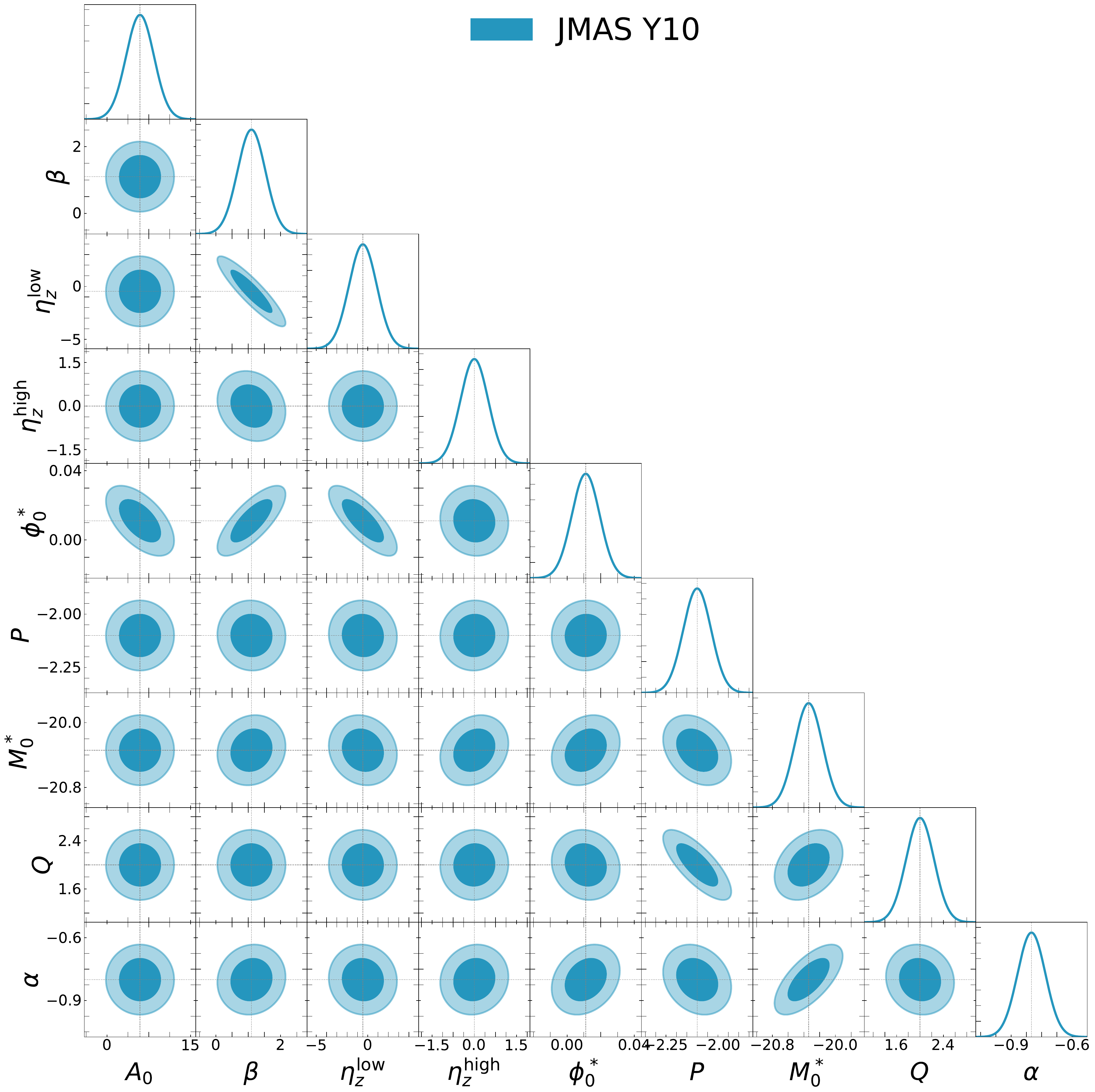}
\caption{IA and LF parameters for LSST Y10.}
\label{fig:ialf_y10}
\end{subfigure}
\caption{\textit{Segments of the Fisher matrix marginalized contours for JMAS analysis across forecasting years.}
The left panel illustrates the correlations between IA and LF parameters for LSST Year 1 (Y1, green contours), while the right panel shows the same for Year 10 (Y10, blue contours).
This side-by-side arrangement highlights the progression of parameter constraints over time.}
\label{fig:corner_ialf_side_by_side}
\end{figure*}

We now examine the interplay and degeneracies between intrinsic alignment (IA) parameters and the luminosity function parameters. 
These correlations are given in Figure \ref{fig:corner_ialf_side_by_side}.

\textbf{Overall intrinsic alignment amplitude} $\boldsymbol{A_0}$.
The observed anti-correlation between $A_0$ and $\phi_0^*$ illustrates the direct relationship between the IA signal strength and the fractional number density of galaxies subject to IA effects (in this work: red galaxies). 
The direct multiplication of $A_0$ and $\phi_0^*$ within the IA model (see Eq. \ref{eq:A_mlim_z}) explains their negative correlation: increasing $\phi_0^*$, and hence the fraction of red galaxies, can be compensated by decreasing the overall IA strength $A_0$. 

$A_0$ shows no significant correlation with $P$, $M_0^*$, $Q$, or $\alpha$ in either forecasting year.
This lack of correlation suggests that the parameters describing primarily the redshift evolution of the luminosity function do not have a direct or observable impact on the overall amplitude of the intrinsic alignment signal. 

\begin{figure*}
\centering
\begin{subfigure}{0.49\textwidth}
\centering
\includegraphics[height=0.35\textheight]{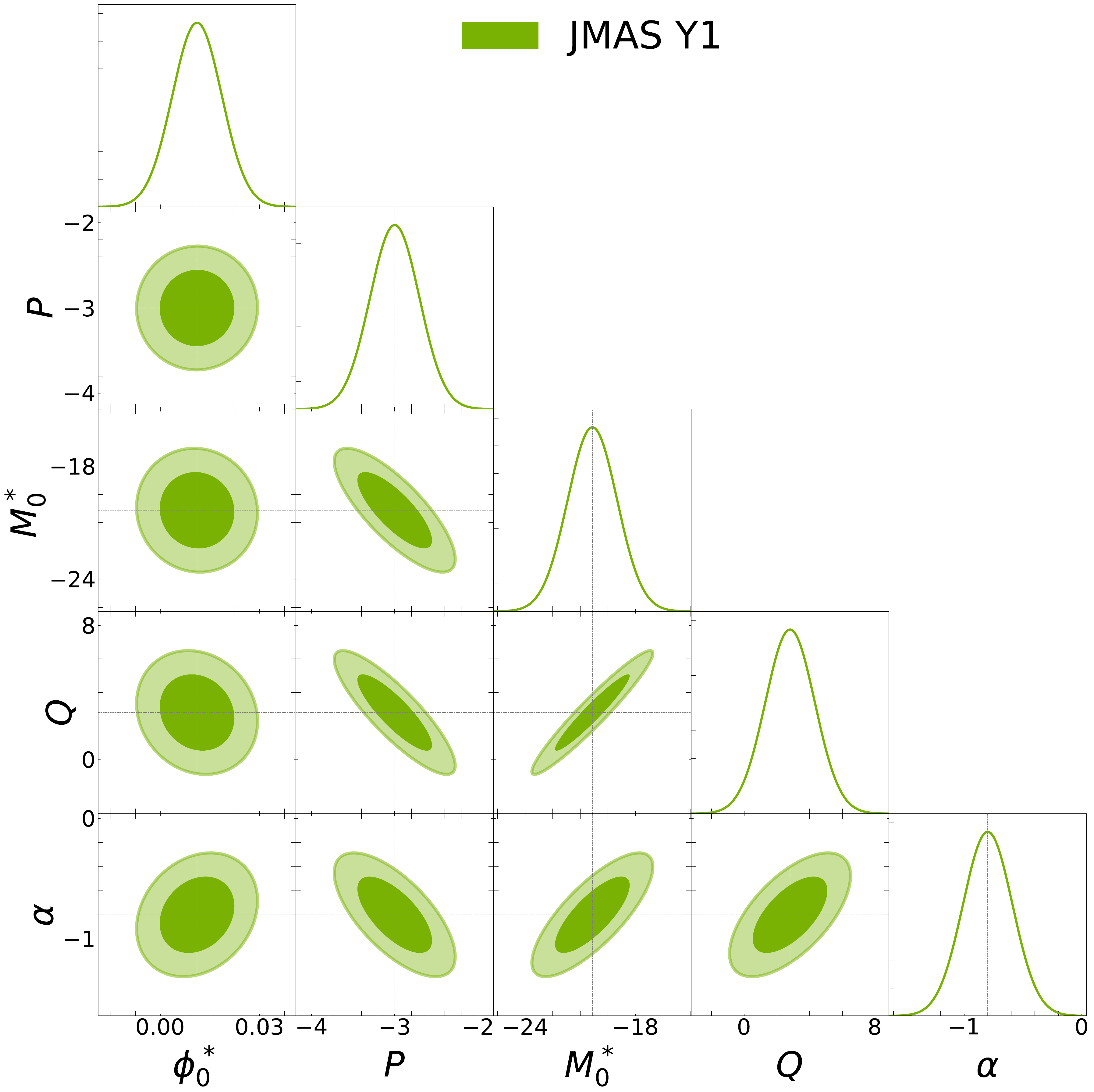}
\caption{LF parameters for LSST Y1.}
\label{fig:lf_y1}
\end{subfigure}
\hfill
\begin{subfigure}{0.49\textwidth}
\centering
\includegraphics[height=0.35\textheight]{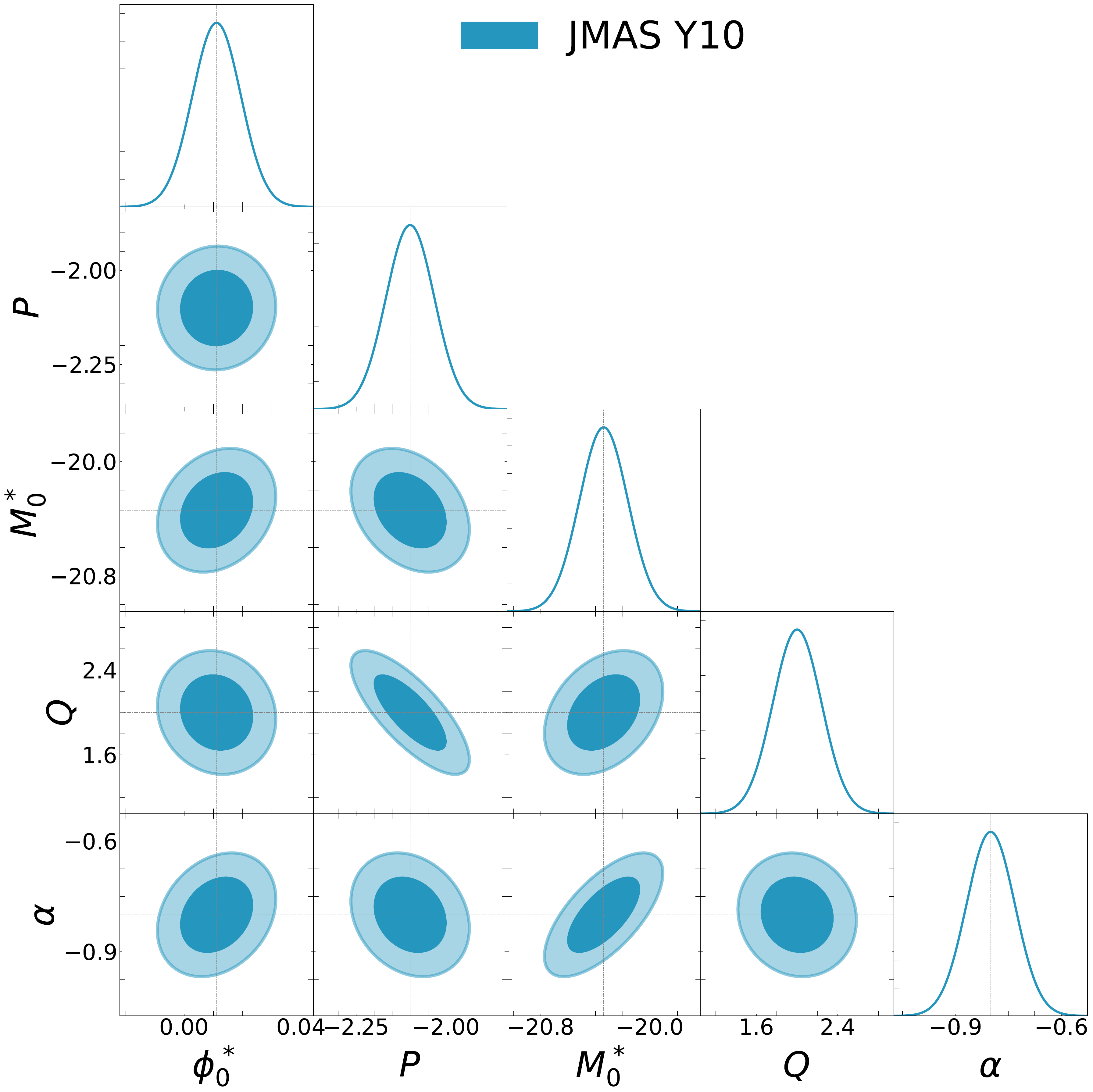}
\caption{LF parameters for LSST Y10.}
\label{fig:ialf_y10}
\end{subfigure}
\caption{\textit{Segments of the Fisher matrix marginalized contours for JMAS analysis across forecasting years.}
The left panel illustrates the correlations between the LF parameters for LSST Year 1 (Y1, green contours), while the right panel shows the same for Year 10 (Y10, blue contours).
This side-by-side arrangement highlights the progression of parameter constraints over time.}
\label{fig:corner_lf_side_by_side}
\end{figure*}

\textbf{Power-law luminosity scaling} $\boldsymbol{\beta}$.
As $\beta$ is the power law index of the luminosity-dependent term in the IA amplitude, we would expect it to be degenerate with the parameters of our luminosity function.
In Year 1, we see this to be generically true. 
However, all but the correlation with $\phi_0^*$ effectively vanish in Year 10, which is quite surprising.
This might indicate a complex interplay between galaxy properties and the intrinsic alignments that evolves with the statistical power of our data, and moreover is potentially influenced by the specific adjustments made to the fiducial LF parameters to match the Y1 and Y10 redshift distribution as appropriate.

\textbf{Low redshift scaling} $\boldsymbol{\eta_z^{\mathrm{low}}}$.
In both Year 1 and Year 10, a clear anti-correlation is observed between $\eta_z^{\text{low}}$ and $\phi_0^*$.
This trend, which tightens in Year 10, indicates that we can compensate the effect of a higher amplitude of the LF function with a smaller explicit redshift dependence in the IA amplitude and vice-versa. 
In Year 1, there is a very slight indication of a positive correlation between $\eta_z^{\text{low}}$ and $P$ (the galaxy density evolution parameter), and a negative correlation with $\eta_z^{\text{low}}$ and $M_0^*$, $Q$, and $\alpha$. 
However, these correlations effectively disappear in Year 10.
The disappearance of these correlations in Year 10 could be attributed again to the different fiducial values of the LF parameters, potentially altering the relationship between galaxy luminosity properties and IA scaling.

\textbf{High redshift scaling} $\boldsymbol{\eta_z^{\mathrm{high}}}$.
In examining the high-redshift scaling factor $\eta_z^{\text{high}}$ of the intrinsic alignment (IA) model across the LSST forecasting years, we observe a notable absence of correlations with the Schechter luminosity function parameters. 
The setting of the central value of $\eta_z^{\text{high}}$ to $0$ is pivotal in our analysis.
This baseline assumption implies a neutral or non-contributory scaling of the IA signal at higher redshifts, making it challenging to discern any significant correlations. 
It is important to consider that, since we are taking the derivative around a fiducial value of $0$, the sensitivity of the Fisher matrix analysis to changes in $\eta_z^{\text{high}}$ might be lower than it would be for a non-zero fiducial value.
This approach could reduce the model's sensitivity to variations in $\eta_z^{\text{high}}$, especially if the influence of this parameter is subtle and requires more pronounced deviations from $0$ to be detectable.
By Year 10, a marginal direct correlation emerges between $\eta_z^{\text{high}}$ and $M_0^*$, hinting at a potential, albeit minimal, influence of the characteristic magnitude of galaxies on the high-redshift scaling of the IA signal.
However, the broader lack of significant correlations, particularly with parameters like $\phi_0^*$, $P$, $Q$, and $\alpha$, might also reflect the limitations of the dataset at high redshifts or the inherent insensitivity of the Fisher matrix to derivatives around a zero fiducial value.

\subsubsection{Degeneracies: within the luminosity function parameter space}\label{subsec:degeneracies_and_biases_lf_vs_lf}
In analyzing the correlations and degeneracies between the Schechter luminosity function parameters themselves, particularly considering their tuning for the LSST Year 1 and Year 10 forecasts, we observe a nuanced interplay.
This parameter space is given in Figure. \ref{fig:corner_lf_side_by_side}.
For $\phi_0^*$, the absence of correlation with $P$ in both Year 1 and Year 10 suggests that the normalization factor of the galaxy number density operates somewhat independently of its evolution parameter.
In fact, $\phi_0^*$ has minimal degeneracy with any of the other luminosity function parameters, which is perhaps unsurprising since it serves only as an overall normalization whereas the others all provide some redshift evolution effects.
Parameter $P$ exhibits a strong anti-correlation with $M_0^*$, $Q$, and $\alpha$ in both years, which can be interpreted as being due to the fact that this set of parameters all induce shifts in the means of the source redshift distributions.
The strong positive correlation of $M_0^*$ with both $Q$ and $\alpha$ in both years, tightening in Year 10, further highlights this effect. 
$Q$ shows an evolving relationship with $\boldsymbol\alpha$, transitioning from a positive correlation in Year 1 to a slightly negative one in Year 10.
We expect that this is due to the different fiducial values of the LF parameters, corresponding to a shift in how the rate of change in galaxy luminosity correlates with the abundance of fainter galaxies over time. 

The complex interplay between luminosity function (LF) parameters, intrinsic alignment (IA) models, and cosmological parameters underscores the need for an even more extensive investigation.
Notably, the parameter space of the LF can be reduced from five to three parameters, as defined in Eq. \ref{eq:lf_params}, streamlining our analysis without sacrificing depth or accuracy. 
Given the intricacies of how these elements influence each other, a comprehensive approach is necessary. 
A Markov Chain Monte Carlo method, with its capability to dispense with simplifying assumptions of Fisher forecasting (e.g., Gaussian likelihoods), presents a promising route.
\section{Summary and Conclusions}
\label{sec:conclusions}

Observational cosmology currently undergoes a data revolution driven by future large area programs like the Rubin Observatory Legacy Survey of Space and Time  \citep{2019ApJ...873..111I}, the Roman Space Telescope \citep{2015arXiv150303757S} and Euclid \citep{2011arXiv1110.3193L}.
Future weak lensing analyses will combine data sources from multiple surveys, including spectroscopic programs like DESI, to fully exploit the cosmological information and calibrate sources of systematics.
This requires the modelling of several complex survey selection functions and their incorporation into the joint analysis.
A consistent modelling of population statistics and other quantities of interest across future multisurvey data vectors is paramount to develop more physical and statistical descriptions of galaxy populations and improve cosmological parameter constraints.
It also allows us to to test for model inconsistencies and model misspecification error.

The presented methodology builds on prior work by \citet{Sheth_2010} that proposed the separate modelling of luminosity function and volume selection in the context of photometric redshift estimation. \citet{van_Daalen_2018} later considered this approach in the context of galaxy clustering redshift estimation and studied the advantages of this approach in the context of the modelling of galaxy-dark matter bias.
We extend and complement the literature by considering a weak lensing data vector and including the luminosity function into the modelling of the intrinsic alignment (IA) signal. 
We detail our approach in \S~\ref{sec:theory_and_methods}. 
We consider the following questions:
\begin{itemize}
    \item \textit{What are the critical parameter degeneracies of our joint IA-photo-z modelling in the context of an LSST-like weak lensing analysis?} 
    \item \textit{What are the implications for the posterior cosmological parameter constraints and the calibration of systematics in the advent of multisurvey analyses? How do our results compare with a traditional approach?}
\end{itemize}
To address these questions \S~\ref{sec:forecasting_setup} presents a Fisher forecast set-up within the context of both LSST Y1 and Y10, to study the effect of this model extension on cosmological parameter constraints.
We illustrate the impact of the luminosity function parameters on the sample redshift distribution and cosmic shear angular power spectra in \S~\ref{sec:lf_params_impact_dndz} and \S~\ref{sec:lf_params_impact_datavector}. In \S~\ref{subsec:results_ia_lf_only} we marginalize over the luminosity function  parameters present in the IA modelling in an otherwise standard analysis approach, and illustrate the high sensitivity of the Fisher information of the intrinsic alignment model parameters on variations of luminosity function parameters.
 Specifically, we see that upon marginalizing over the luminosity function parameters the $1\sigma$ constraints on dark energy equation of state parameters degrade by 20\%, while the IA parameter constraints are affected up to the 200\% level for forecasting year 10.
This exploration highlights the importance and possibilities presented by the complete incorporation of luminosity function parameters into cosmological investigations.

In Section \S~\ref{subsec:jmas_results}, we compare forecast cosmological results within our full joint analysis framework as compared to a standard analysis approach, and discuss the degeneracies that emerge among cosmological parameters, intrinsic alignment (IA) parameters, and luminosity function (LF) parameters. We find that our approach can produce cosmological parameter constraints which are broadly comparable to that of a standard approach while allowing for greater physical interpretability. We do observe a modest decrease in the precision of some cosmological parameter constraints when we incorporate the modelling of a luminosity function into our framework.
However, this slight reduction in precision is arguably compensated for by the significant advantages gained through a more comprehensive and physically motivated modelling approach. 

Notably, we uncover interdependencies within the LF parameter sets and observe their interactions with the broader set of cosmological and IA parameters. 
Luminosity function parameters, $\phi_0^*$, $P$, $M_0^*$, $Q$, and $\alpha$, have a direct impact on the abundance and properties of galaxies as a function of redshift.
Consequently, variations in these LF parameters can influence the observed redshift distribution of source galaxies, which in turn affects cosmic shear measurements.
This connection between LF parameters and cosmological parameters underscores the value of introducing additional constraints on the luminosity function to enhance our understanding of these relationships. 

Looking ahead, the framework presented in this work can be extended to encompass a joint treatment of sample redshift distribution functions, galaxy-dark matter bias, and intrinsic alignments in the context of a full 3x2pt analysis.
The proof-of-concept offered here also paves the way for application of a similar approach to the treatment of selection function, including with respect to luminosity in redshift, in the more direct estimation of redshift distributions and their joint inference with IA and galaxy bias.
These future endeavors highlight the versatility and promise of joint modelling frameworks such as the one presented here, which have the potential to significantly enhance our understanding of the universe and its underlying cosmological parameters. 
\section*{Acknowledgements}
This paper has undergone a DESC internal review. 
We are deeply grateful to Elisa Chisari and Shahab Joudaki for their insightful contributions during the DESC internal review process.
NS and CDL acknowledge funding from a 2021 LSSTC Science Catayst (formerly `Enabling Science' grant).
NS and CDL thank the organisers of the Lorentz Centre Workshop hol-IA as well as the follow up working meeting of the DESC-IA topical team at the University of Utrecht for the opportunity for productive discussions on the topic of this work.
NS thanks H.  Almoubayyed, M. Bonici, E. Chisari, K. Cranmer, N. Cross, D. Hogg, R. Jimenez, R. Mandelbaum,  K. Morå, and M. van der Wild for useful discussions.
The DESC acknowledges ongoing support from the Institut National de Physique Nucl\'eaire et de Physique des Particules in France; the Science \& Technology Facilities Council in the United Kingdom; and theDepartment of Energy, the National Science Foundation, and the LSST Corporation in the United States.
DESC uses resources of the IN2P3 Computing Center (CC-IN2P3--Lyon/Villeurbanne - France) funded by the Centre National de la Recherche Scientifique; the National Energy Research Scientific Computing Center, a DOE Office of Science User Facility supported by the Office of Science of the U.S.\ Department of Energy under Contract No.\ DE-AC02-05CH11231; STFC DiRAC HPC Facilities, funded by UK BEIS National E-infrastructure capital grants; and the UK particle physics grid, supported by the GridPP Collaboration.
This work was performed in part under DOE Contract DE-AC02-76SF00515.

\textbf{Software Acknowledgments.} 
The success of this research is largely attributable to several key software tools and packages. 
Foremost, we acknowledge the foundational role of \pkg{Python}\footnote{\url{https://www.python.org}}. 
Our analysis heavily relied on essential \pkg{Python} libraries including \pkg{Numpy} \citep{numpy}, \pkg{Scipy} \citep{scipy}, \pkg{Matplotlib} \citep{matplotlib}, \pkg{HDF5} \citep{hdf5}, and \pkg{Pandas} \citep{pandas}. 
The generation of corner plots was facilitated by the \pkg{GetDist} package \citep{lewis2019getdist}, while the \pkg{CMasher} package \citep{cmasher_vandervelden} provided invaluable colormaps enhancing our results' visualization.
Finally, the cosmological analysis "heavy lifting" was adeptly supported by the \pkg{CCL} \citep{Chisari_2019}, for which we extend our deepest gratitude to the LSST DESC CCL team.

\textbf{Author contributions.}
NS wrote code, performed analysis, and contributed extensively to paper writing.
DL and MR: initialised project idea, provided scientific guidance and supervision throughout project, contributed to paper writing.
\section*{Data Availability}

Data products from this paper (redshift distributions, redshift dependent IA amplitudes, data vectors, Fisher matrices) can be found on \url{https://zenodo.org/records/11447672}.
Plotting scripts can be found at \url{https://github.com/nikosarcevic/JMAS}.
 

\bibliographystyle{mnras}
\bibliography{joint_modelling}



\appendix
\section{Additional corner plots and tables}

\begin{figure*}
\includegraphics[width=0.8\textwidth]
{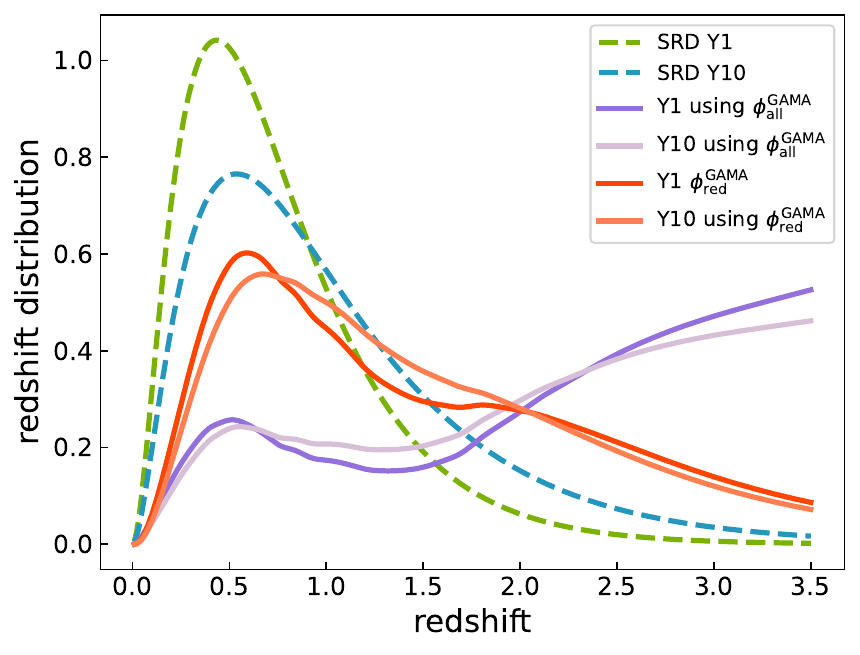}
\caption{\textit{Comparing SRD and redshift distributions obtained using luminosity functions from GAMA survey.}
In this figure, we present the LSST Science Requirements Document (SRD) redshift distributions for Y1 and Y10 – represented by green and blue dashed lines, respectively – with distributions derived from a convolution method using luminosity functions (LFs) from the GAMA survey \citep{Loveday_2011}.
Specifically, we apply the luminosity function parameters for $r$-band red and all galaxy types as detailed in the intrinsic alignment model by \citet{Krause_2015}.
The comparison underscores that the GAMA-based parameters, when used directly, fail to replicate the desired LSST SRD distributions.
This discrepancy highlights the necessity to deviate from the GAMA LFs parameters to achieve congruence with the LSST SRD projections.
}
\label{fig:nz_w_gama_lfs}
\end{figure*}

\begin{figure*}
\includegraphics[width=0.8\textwidth]
{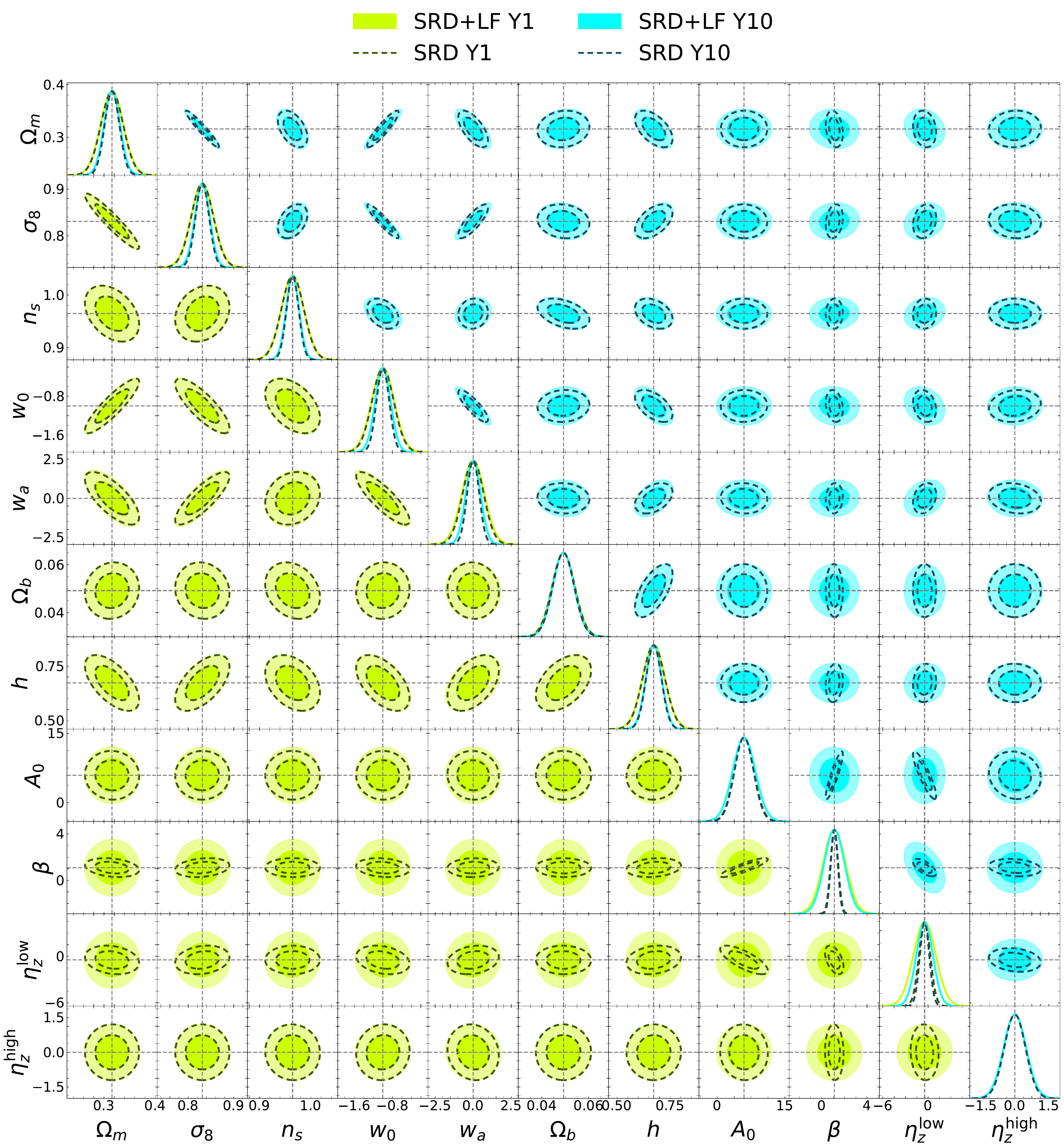}
\caption{
\textit{Comparative analysis between forecasts with (SRD+LF) and without (SRD) luminosity function (LF) parameters.}
This corner plot illustrates the correlations and distributions of seven cosmological parameters and four intrinsic alignment (IA) parameters within the framework of the LSST cosmic shear analysis. 
It is divided into two triangular sections, representing forecasts for different years and scenarios.
The SRD (Science Requirements Document) scenario, a standard analysis setup \citep{srd_lsst_desc}, utilizes a fixed Schechter luminosity function derived from GAMA and DEEP2 surveys without varying its parameters in IA modeling. 
In contrast, the SRD+LF (SRD plus Luminosity Function) scenario, as suggested in \citet{Krause_2015}, expands upon this by varying luminosity function parameters within the IA modeling, providing a more dynamic and potentially more accurate representation of cosmic shear effects.
In the lower triangle of the plot, the Year 1 (Y1) forecast is depicted. 
Here, the SRD+LF scenario is represented by filled chartreuse ellipses, highlighting the parameter space and correlations with varied LF parameters.
Overlaid on these are unfilled khaki green contours for the SRD scenario, offering a visual comparative analysis. 
The filled contours indicate tighter parameter constraints under the SRD+LF scenario, in line with the symmetric percentage differences observed, particularly for parameters like $\Omega_m$, $\sigma_8$, and the scale factor $n_s$ which show modest deviations in Y1.
The upper triangle displays the Year 10 (Y10) forecast.
Filled neon blue ellipses delineate the SRD+LF scenario, with unfilled denim blue contours for SRD. 
This section starkly demonstrates the increasing divergence between the two scenarios over time, especially for parameters like  $w_0$ and $w_a$, where the SRD+LF scenario reveals significantly tighter constraints.
This is evident in the symmetric percentage differences exceeding $20\%$ for these parameters in the Y10 forecast.
Along the diagonal, Gaussian distributions for each parameter indicate the precision of parameter estimates.
}
\label{fig:srd_vs_srd+lf_contours}
\end{figure*}


\begin{figure*}
    \centering
\includegraphics[width=\textwidth]{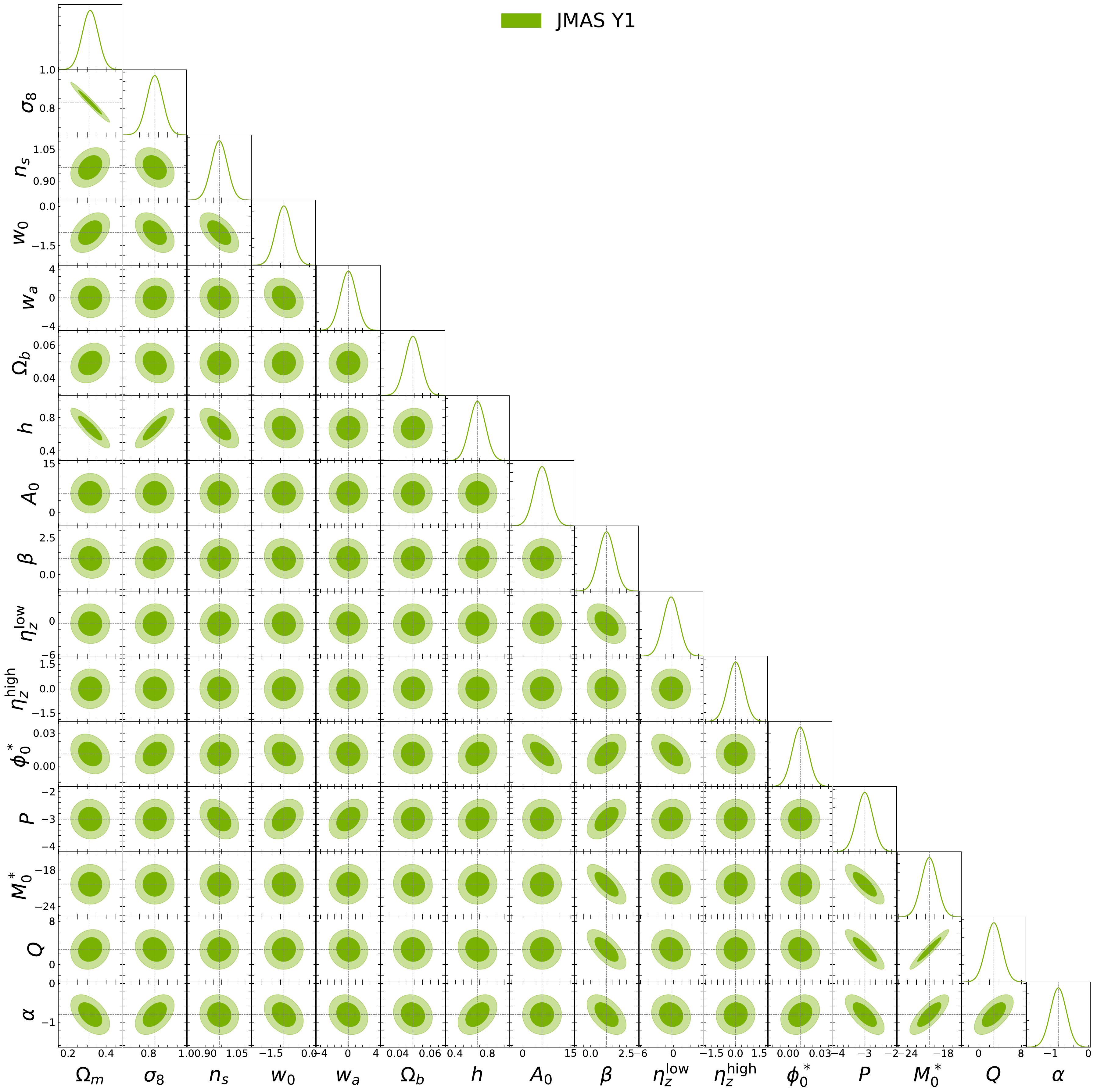}
 \caption{\textit{Fisher matrix marginalized contours for the LSST Y1 joint modelling analysis}
}
\label{fig:corner_jmas_y1_cosmoialf}
\end{figure*}

\begin{figure*}
    \centering
\includegraphics[width=\textwidth]{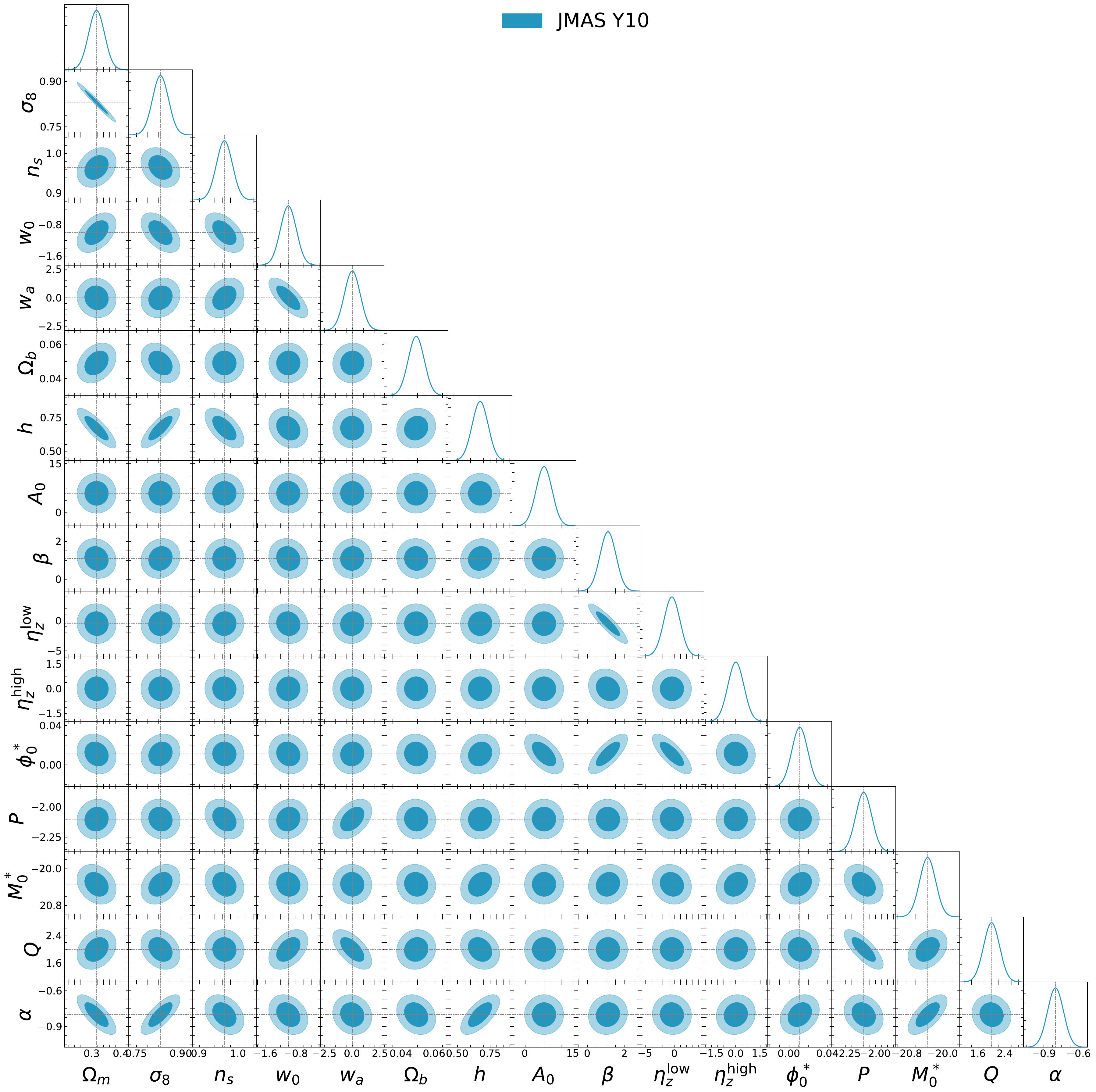}
 \caption{\textit{Fisher matrix marginalized contours for the LSST Y10 joint modelling analysis.}  
}
\label{fig:corner_jmas_y10_cosmoialf}
\end{figure*}

\begin{table*}
\centering
\caption{\textit{Marginalized $1\sigma$ errors on cosmological and intrinsic alignment parameters for different forecasting scenarios explored in this work. }
The table presents a comparative analysis between the constraining powers of SRD, SRD+LF, and JMAS across two forecast periods: LSST year 1 and year 10. 
The upper block of the table displays results for the LSST year 1 forecast, highlighting how each framework influences the precision of parameter estimation. 
The bottom segment focuses on year 10, illustrating the potential evolution in constraining power over time.
This comparison sheds light on the effectiveness of incorporating JMAS methodologies in exploring the feasibility of cosmological forecasts.
}
\label{tab:srd_srdlf_jmas_stds}
\begin{tabular}{cccccccccccc}
\hline
Forecast & \multicolumn{11}{c}{\textbf{Parameter}} \\ \hline
& $\Omega_{m}$ & $\sigma_{8}$ & $n_s$ & $w_{0}$ & $w_{a}$ & $\Omega_{b}$ & $h$ & $A_0$ & $\beta$ & $\eta_{z}^\mathrm{low}$ & $\eta_{z}^\mathrm{high}$ \\ \hline
\multicolumn{12}{c}{\textbf{Year 1}} \\ \hline
SRD & 0.022 & 0.024 & 0.022 & 0.229 & 0.688 & 0.005 & 0.052 & 2.156 & 0.334 & 0.754 & 0.493 \\
SRD+LF & 0.022 & 0.025 & 0.022 & 0.237 & 0.739 & 0.005 & 0.053 & 2.500 & 0.998 & 1.490 & 0.497 \\
JMAS & 0.042 & 0.042 & 0.038 & 0.310 & 1.146 & 0.005 & 0.098 & 2.500 & 0.554 & 1.440 & 0.498 \\
\hline
\multicolumn{12}{c}{\textbf{Year 10}} \\ \hline
SRD & 0.014 & 0.015 & 0.012 & 0.134 & 0.397 & 0.005 & 0.036 & 2.103 & 0.307 & 0.622 & 0.469 \\
SRD+LF & 0.016 & 0.016 & 0.013 & 0.162 & 0.485 & 0.005 & 0.037 & 2.500 & 0.896 & 1.101 & 0.499 \\
JMAS & 0.027 & 0.027 & 0.021 & 0.208 & 0.714 & 0.005 & 0.061 & 2.500 & 0.430 & 1.354 & 0.494 \\
\hline
\end{tabular}
\end{table*}


\begin{figure*}
    \centering
\includegraphics[width=\textwidth]{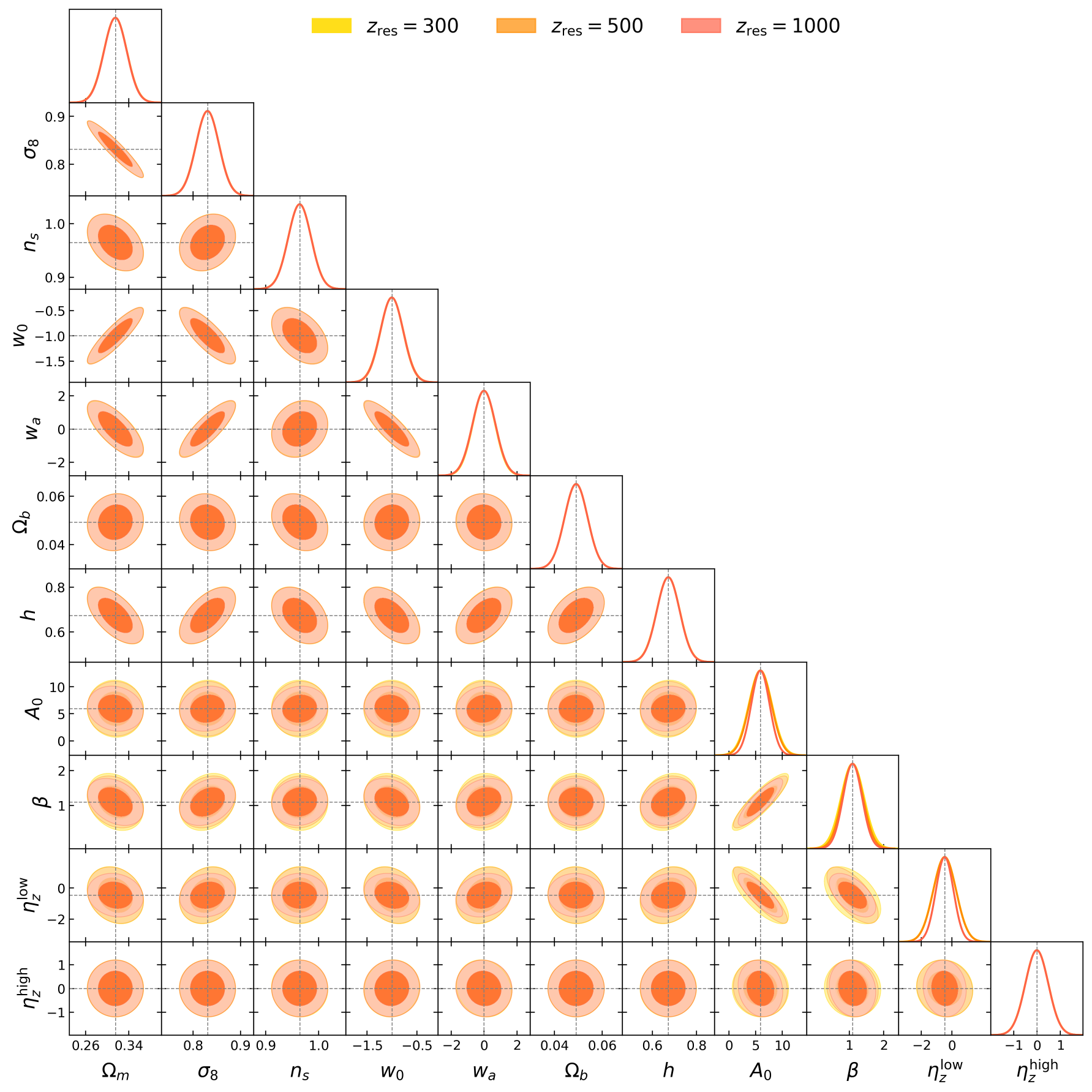}
 \caption{\textit{Stability Test of Fisher Matrix Marginalized Contours for Various Redshift Resolutions.} 
 This figure compares the stability of Fisher matrix marginalized contours for three forecasting scenarios, all based on the SRD Y1 model.
 The contours in yellow represent a forecast with redshift range resolution of 300, followed by resolutions of 500 (in orange) and 1000 (in tomato).
 Notably, when using 'stem' derivatives, the redshift resolution does not significantly impact the stability of Fisher contours, which contrasts with the behavior observed when employing finite difference methods for derivatives (e.g., 5-point stencil or \pkg{numdifftools} library).
 Only minor differences are observed in the intrinsic parameters due to large errors and uncertainties in these parameters, while the Fisher matrices remain remarkably stable against numerical variations.}
\label{fig:corner_srd_stem_test_z_res}
\end{figure*}

\begin{figure*}
    \centering
\includegraphics[width=\textwidth]{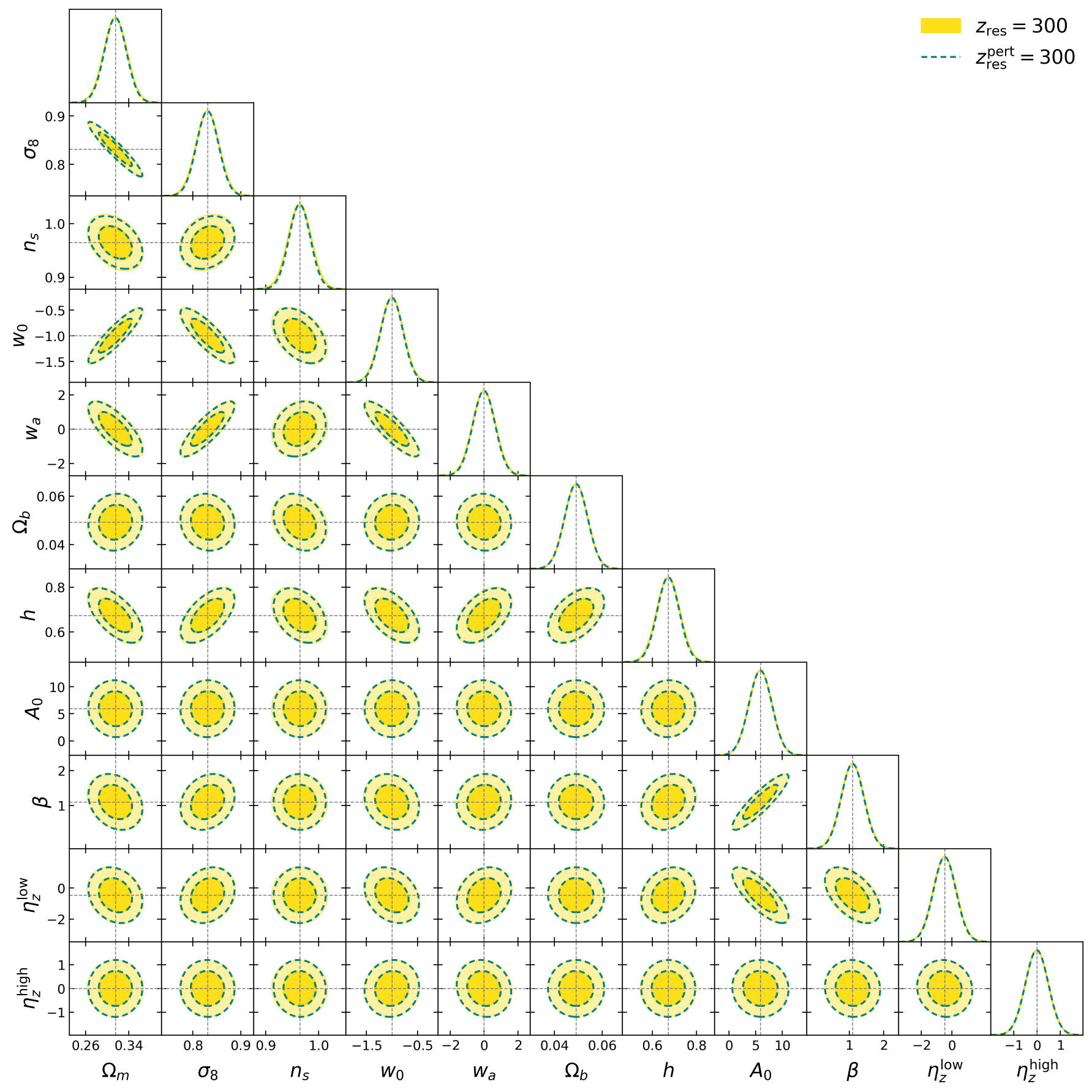}
 \caption{\textit{Stability Test of Perturbed Derivatives.}
 This figure presents a comparison between two SRD Y1 forecasts with a redshift range resolution of 300, one with unperturbed derivatives (yellow filled contours) and another with derivatives perturbed by 10\% (unfilled dashed teal contours).
 Remarkably, no significant variations are noticeable in the Fisher matrix marginalized contours, highlighting the robust stability of the forecasts with stem derivatives routine, even with perturbed derivatives."}
\label{fig:corner_srd_stem_test_z_res_pert}
\end{figure*}

\bsp	
\label{lastpage}
\end{document}